\documentclass{ws-ijmpd}
\usepackage[super,compress]{cite}
\begin{document}

\markboth{Capovilla, Cruz, Rojas}
{Ostrogradsky-Hamilton approach to geodetic brane gravity}

%%%%%%%%%%%%%%%%%%%%% Publisher's Area please ignore %%%%%%%%%%%%%%%
%
\catchline{}{}{}{}{}
%
%%%%%%%%%%%%%%%%%%%%%%%%%%%%%%%%%%%%%%%%%%%%%%%%%%%%%%%%%%%%%%%%%%%%

%%%%%%%%%%%%%%%%% SOME DEFINITIONS  %%%%%%%%%%%%%%

\newcommand{\D}{\mathcal{D}}
\newcommand{\Dt}{\widetilde{\mathcal{D}}}
\newcommand{\be}{\begin{equation}}
\newcommand{\ee}{\end{equation}}
\newcommand{\beq}{\begin{eqnarray}}
\newcommand{\eeq}{\end{eqnarray}}
\newcommand{\sk}{\smallskip\noindent}
\newcommand{\bk}{\bigskip\noindent}
\newcommand{\n}{\nonumber}
%%%%%%%%%%%%%%%%%%%%%%%%%%%%%%%%%%%%%%%%%%%%%%%

%%%%%%%%%%%%%%%%%%%%%%%%%%%%%%%%%%%%%%%%%%%%%%%%%%%%%%%%%%%%%%%%
\title{OSTROGRADSKY-HAMILTON APPROACH TO GEODETIC BRANE GRAVITY
%\footnote{For the title, try not to use more than 3 lines.
%Typeset the title in 10~pt Times roman, uppercase and boldface.}  
}

\author{RICCARDO CAPOVILLA%\footnote{Typeset names in
%8~pt roman, uppercase. Use the footnote to indicate the
%present or permanent address of the author.}
}

\address{\textit{Departamento de F\'\i sica, CINVESTAV-IPN, Av. 
Instituto Polit\'ecnico Nacional 2508, 
\\
Col. San Pedro Zacatenco, 07360, Gustavo A. Madero, 
CDMX, M\'exico
\\
capo@fis.cinvestav.mx}}

\author{GIOVANY CRUZ}

\address{\textit{Departamento de F\'\i sica, CINVESTAV-IPN, Av. 
Instituto Polit\'ecnico Nacional 2508, 
\\
Col. San Pedro Zacatenco, 07360, Gustavo A. Madero, 
CDMX, M\'exico
\\
gcruz@fis.cinvestav.mx}}

\author{EFRA\'IN ROJAS}

\address{\textit{Facultad de F\'\i sica, Universidad 
Veracruzana, Paseo No. 112, Desarrollo Habitacional,
\\
Nuevo Xalapa, 91097, Xalapa-Enr\'\i quez, Veracruz, 
M\'exico
\\
efrojas@uv.mx}}

\maketitle

\begin{history}
\received{Day Month Year}
\revised{Day Month Year}
\end{history}

\begin{abstract}
We develop the Ostrogradsky-Hamilton formalism for  geodetic 
brane gravity, described by the Regge-Teitelboim geometric 
model in higher codimension. We treat this  gravity theory 
as a second-order derivative theory,  based on the extrinsic 
geometric structure of the model. As opposed to previous 
treatments of geodetic brane gravity,  our Lagrangian is 
linearly dependent on second-order time derivatives of the 
field variables, the embedding functions. The difference resides 
in a  boundary term in the action, usually discarded. Certainly, 
this suggests applying an appropriate Ostrogradsky-Hamiltonian 
approach to this type of theories. The price to pay for this 
choice is the appearance of second class constraints. We 
determine the full set of phase space constraints, as well as 
the gauge transformations they generate in the reduced phase 
space. Additionally, we compute the algebra of constraints and 
explain its physical content. In the same spirit, we deduce 
the counting of the physical degrees of freedom. We comment 
briefly on the naive formal canonical quantization emerging 
from our development.
\end{abstract}

\keywords{Ostrogradsky-Hamilton framework; Regge-Teitelboim model; Variational techniques.}

\ccode{PACS numbers: 04.50.+h, 04.20.Fy, 11.25.-w}

%\tableofcontents

%%%%%%%%%%%%%%%%%%%%%%%%%%%%%%%%%%%%%%%%%%%%%%%%%
\section{Introduction}
\label{sec1}
%%%%%%%%%%%%%%%%%%%%%%%%%%%%%%%%%%%%%%%%%%%%%%%%%%

In 1975, T. Regge and C. Teitelboim  [RT] pictured 
our four-dimensional spacetime as the trajectory of a three-dimensional extended 
object evolving geodesically in a fixed higher-dimensional background Minkowski spacetime, close in  spirit to the worldline/worldsheet 
behavior of classical relativistic objects, like strings~\cite{RT1975}.
They were motivated initially by a search for an alternative theory 
to pursue quantum gravity, rather than continue to quantize 
pure gravity described by the Einstein-Hilbert [EH] action 
in terms of a metric field variable, considering the unsurmountable difficulties encountered in a standard perturbative approach~\cite{veltman1976}. In the following we will refer to standard General Relativity based on the EH action as metric-GR,
see~\cite{wald}. The RT  theoretical viewpoint replaces the components 
of the spacetime metric  by the embedding functions of the worldvolume spanned by the three-dimensional object as field 
variables, so that the worldvolume metric becomes  a composed
field variable. The pioneering work of RT has received renewed interest in the context of braneworld 
scenarios~\cite{arkani1998,randall99,maartens2010}, and in particular through the studies of Davidson and collaborators~\cite{davidson1998,davidson2003}, that introduced the
term `Geodetic Brane Gravity' [GBG].

On geometrical grounds, in order to ensure the local existence 
of an embedding framework, at most $N = n(n+1)/2$ dimensions 
are needed for the ambient spacetime background. In addition, it is known that if the worldvolume metric admits Killing vectors, that number can be
reduced~\cite{janet1926,cartan1927,friedman1961}. 
The RT model consists  of the integral over the worldvolume of the Ricci scalar constructed from the induced worldvolume metric, so that one may have a misconception of continuing to work with metric-GR because the action is the same. However, one must keep in mind that in this framework the field variables are given by the embedding functions. Either way, the equations of motion remain second-order in derivatives of the 
fields, as in GR \cite{maia1986,pavsic1986,tapia1989}. 
In both cases, GBG and metric-GR, the action presents a problem due to the appearance of a term linear in the
second derivative of the field variables. 

After the RT proposal came up, the idea was criticized due to 
gauge dependance~\cite{Deser1976} and, as Regge and Teitelboim themselves stated in their work, for the lack of an appropriate Hamiltonian formulation. Regarding the latter, many authors 
have made important advances in this direction, using several 
Hamiltonian strategies
\cite{davidson1998,davidson2003,capo2006,ostro2009,paston2007,paston2010}. Since the RT model is an embedded geometric model 
in terms of the  geometry of sub-manifolds, where a 
divergence term can also be identified, perhaps avoiding 
that term as it is generally done may be detrimental to 
obtaining a clear canonical formulation.

In this paper we reconsider the canonical formalism of the RT 
model by making use of the appropriate Ostrogradsky-Hamilton\cite{Ostro} framework developed for singular systems. Our 
approach has the advantage of bringing to the forefront the full 
geometrical content of the model in any  codimension as well to consider the effect of all geometric terms in the RT model. In fact, we borrow the existent 
Hamiltonian formulation for relativistic extended objects \cite{ADM-2000,ham-branes2004} that, in turn, was inspired by the Arnowitt-Deser-Misner Hamiltonian formulation of GR [ADM]. The canonical analysis does not involve reduction 
of the RT model by eliminating a total divergence. The resulting Lagrangian is simply linear in the accelerations of the extended 
object, that make this, in principle, a second-order derivative 
theory; fortunately, the accelerations enter into the game in 
such a way that they do not raise the order of the equations of motion. 
Under this consideration and, according to the Ostrogradsky line of 
reasoning, the canonical approach needs to double the number of 
phase space variables. The advantage of this treatment is that we 
manage to keep the original geometric nature of the model intact. Additionally, 
this fact makes more evident the role that both the momenta and 
the Hamiltonian constraints play within the canonical structure. 
To the best of our knowledge, this is the first attempt 
in this direction, at least for the full theory. 

After obtaining the full set of constraints 
in phase space, we separate them into first and second-class 
constraints. In particular, we compute the Dirac algebra of 
the constraints and analyze its content in addition to study 
the gauge transformations that the first-class constraints 
generate. Nevertheless, it should be mentioned that this point of view has 
been adopted earlier in an Hamiltonian formulation of the RT 
model minisuperspace by Cordero et al. \cite{ostro2009}, 
and then by Banerjee et al \cite{BMP}, where the field 
theory is reduced to a finite number of degrees of freedom, 
allowing to substantial progress in the quantization of the 
reduced model.
In passing, we would like to mention that, along 
this line of reasoning, Dutt and Dresden were interested in try to
apply the Ostrogradsky-Hamilton formulation to 
metric-GR~\cite{dutt1986}.

The paper is organized as follows. In Sec. \ref{sec2} we introduce 
our notation and provide an overview of pure GBG, without additional 
brane matter fields, via the  RT geometric model.  
In Sec. \ref{sec3}, we consider a $p+1$ ADM decomposition of the 
worldvolume geometry and obtain a suitable ADM Lagrangian for GBG 
in terms of ADM field variables, linear in the acceleration. 
The Ostrogradsky-Hamilton formulation is the subject of 
Sec \ref{sec4}, where we construct the phase space appropriate for a 
higher derivative theory, the Hamiltonian  and we identify  the primary 
and secondary phase space constraints for the theory. We also calculate Hamilton's equation as a check for consistency. In Sec. \ref{sec5}, 
we separate the primary and secondary phase space constraints in 
first and second class constraints, and compute their algebra, using 
the standard Dirac-Bergmann approach. We also consider infinitesimal 
canonical transformations, as an help towards understanding the
meaning of the constraints themselves. We end in Sec. \ref{sec6} 
with a brief discussion. In  three Appendices we have collected useful 
results used throughout the main text.

%%%%%%%%%%%%%%%%%%%%%%%%%%%%%%%%%%%%%%%%%%%%%%%%%%%%%%%%%%%%%%%%%
\section{Geodetic brane gravity}
\label{sec2}
%%%%%%%%%%%%%%%%%%%%%%%%%%%%%%%%%%%%%%%%%%%%%%%%%%%%%%%%%

We consider a $(p+1)$-dimensional worldvolume $m$ spanned by 
the evolution of a $p$-dimensional spacelike extended object, 
or brane, $\Sigma$ in a flat $N$-dimensional Minkowski 
background spacetime, $\{\mathcal{M}, \eta_{\mu \nu} \}$, with metric $\eta_{\mu\nu} 
= \text{diag} (-1,1,1,\ldots,1)$ ($\mu,\nu = 0,1,2, \ldots,N-1$). 
$m$ is described by the embedding $y^\mu = X^\mu(x^a)$, where 
$y^\mu$ are local coordinates for $\mathcal{M}$, $x^a$ are 
local coordinates for $m$, and $X^\mu$ are the embedding 
functions ($a,b = 0,1,2,\ldots,p$). The vectors $e^\mu{}_a := \partial_a X^\mu$ form a basis of tangent vectors 
to $m$. The inner product among them produce the components 
of the induced metric $g_{ab} = \eta_{\mu\nu} e^\mu{}_a e^\nu{}_b 
= e_a \cdot e_b$, that in this sense is a composed field variable. Here and henceforth a dot  denotes 
inner product using the background Minkowski metric. By 
$g^{ab}$ we denote the inverse of $g_{ab}$, and by $g$ its 
determinant. The worldvolume $m$ is assumed to be timelike so $g < 0$. In this framework, we 
introduce $N- p- 1$ normal vectors to the worldvolume $m$, denoted by $n^{\mu\,i}$, and 
defined implicitly by $n^i \cdot e_a = 0$ and $n^i \cdot n^j = 
\delta^{ij}$ ($i,j = 1,2,\ldots, N - p -1$), up to a sign  and 
a rotation. This gauge freedom  does require the introduction of a gauge field, the twist potential, as shown below .
 We also introduce 
the extrinsic curvature for  $m$ with  $K_{ab}^i = - n^i \cdot \nabla_a 
e_b$ and the mean extrinsic curvature as its trace $K^i = g^{ab} 
K_{ab}^i$ where $\nabla_a$ denotes the torsion-less metric compatible
worldvolume covariant 
derivative, $\nabla_a g_{bc}= 0$. In this spirit, the worldvolume 
scalar curvature depends explicitly on the extrinsic curvature 
by using the contracted  Gauss-Codazzi integrability condition for surfaces 
$\mathcal{R} = K^i K_i - K_{ab}^i K^{ab}_i$~\cite{spivak1970,capovilla1995}.

\smallskip
\noindent
Geodetic brane gravity is described by the  RT model 
defined by~\cite{RT1975}
\be 
S_{\text{\tiny RT}} [X^\mu] = \frac{\alpha}{2} \int_m d^{p+1}x\,
\sqrt{-g}\,\mathcal{R},
\label{action0}
\ee
where $\alpha$ is a constant with dimensions $[L]^{(1-p)}$ in 
natural units. 
We could  add some matter contribution through a Lagrangian 
$L_{\text{\tiny matt}}$ of fields living on the  brane. In any 
case, such a term would not affect the geometric arguments in this note, that focus on the curvature contribution. 
As we are not coupling to any brane matter fields, we set $\alpha = 1$.
Carrying out the first variation of the action we are able to obtain the equations of motion [eom]~\cite{capovilla1995}. The classical brane trajectories are 
obtained from the $N-p-1$ compact relations
\be 
G^{ab} K_{ab}^i = 0,
\label{eom0}
\ee
where $G_{ab} = \mathcal{R}_{ab} - (1/2) \mathcal{R} g_{ab}$ is the 
worldvolume Einstein tensor. These eom are of second order in 
derivatives of the field variables $X^\mu$ because of the presence 
of the extrinsic curvature. Additionally, there are $p+1$ tangential 
vanishing expressions related to the eom, reflecting the reparametrization 
invariance of the action~(\ref{action0}). Indeed, these are given
by the divergence-free condition $\nabla_a G^{ab} = 0$.
Another way of expressing the eom is by using the definition of the extrinsic curvature. In addition, the eom~(\ref{eom0}) can also be 
written as a set of projected conservation laws
\be 
(\nabla_a \mathcal{P}^a) \cdot n^i = 0,
\label{eq3}
\ee
where the conserved stress tensor, $\mathcal{P}_\mu{}^a$, is given by ~\cite{capovilla2000} 
\be 
\mathcal{P}_\mu{}^a := - \sqrt{-g} \,G^{ab} e_{\mu\,b}.
\label{stress}
\ee
Notice that $\mathcal{P}_\mu{}^{a}$ is purely tangential. In fact, 
this feature characterize theories leading to second-order 
eom. Indeed, (\ref{stress}) belongs to a family of conserved
stress tensors associated to second-order derivative
geometrical models leading to second-order eom,  called Lovelock branes ~\cite{bil-2013}.

With an eye towards the Hamiltonian framework, when $\Sigma$ is 
viewed as a spacelike manifold immersed into $m$ (see \ref{app0}), 
the associated timelike unit normal, $\eta^a$, helps to construct 
the linear momentum density on $\Sigma$ with
\be 
\pi_\mu := N^{-1} \eta_a \mathcal{P}_\mu{}^a,
\label{pi}
\ee
where $N$ represents the lapse function that appears in the ADM 
decomposition for geometric extended object models depending 
on the extrinsic curvature, as we will see shortly.

%%%%%%%%%%%%%%%%%%%%%%%%%%%%%%%%%%%%%%%%%%%%%%%%%%%%%%%%%%
\section{The ADM Lagrangian for geodetic brane gravity}
\label{sec3}
%%%%%%%%%%%%%%%%%%%%%%%%%%%%%%%%%%%%%%%%%%%%%%%%%%%%%%%%%%%%%

An adaptation of the ADM framework for metric-GR  to branes, needed for a canonical formulation of GBG, 
is  given in detail in~\cite{ADM-2000,ham-branes2004}. 
If we assume that $m$ is globally hyperbolic, then it is possible
to foliate it into a set of spacelike hyper-surfaces $\Sigma_t$. 
This suggests to split in space and time derivatives the 
relevant  worldvolume geometric quantities, close in  spirit to the ADM 
formulation of metric-GR. 

We describe $\Sigma_t$  using an embedding 
formulation. First, using the embedding $y^\mu=X^\mu (t = 
\text{const}, u^A)$, we split the $p+1$ worldvolume coordinates 
$x^a$ into an arbitrary time parameter $t$ and $p$ coordinates 
$u^A$ with $(A,B = 1,2\ldots,p)$., for $\Sigma_t$.  In this sense, $\Sigma_t$ 
is viewed as the spacelike extended object $\Sigma$ at fixed $t$.
Secondly, it can be described also  by its embedding in $m$ itself, 
$x^a = X^a (u^A)$. Both descriptions are related by composition. 
Indeed, in one  picture, the tangent vectors 
to $\Sigma_t$ are  $\epsilon^\mu{}_A = X^\mu{}_A = 
\partial X^\mu /\partial u^A$, and  then the induced metric on 
$\Sigma_t$ is $h_{AB} =  X_A  \cdot X_B$. 
On the other hand, the tangent vectors to $\Sigma_t$ are 
$\epsilon^a{}_A = X^a{}_A = \partial X^a /\partial u^A$ and 
the induced metric is  $h_{AB} = g_{ab} X^a{}_A X^b{}_B$. 
Notice that $h_{AB} = X_A  \cdot X_B =  
(X_a  \cdot X_b) X^a{}_A X^b{}_B$, and we see that 
$\epsilon^\mu{}_A = e^\mu{}_a \epsilon^a{}_A$, from composition. 
Accordingly, the choice of the hypersurface vector basis depends on the particular 
description we are interested in. For the first description we have 
$\{ \epsilon^\mu{}_A, n^\mu{}_i, \eta^\mu\}$, whereas for the second 
one we have $\{ \epsilon^a{}_A, \eta^a\}$, where the appearance
of the unit timelike vector accounts for the causal structure
on $\Sigma_t$. Note that $\eta^\mu$ is defined
implicitly by $\epsilon_A \cdot \eta = 0$,
$n_i  \cdot \eta  = 0$ and $
\eta \cdot \eta  = -1$, and in the second description we have 
a single unit timelike normal vector, $\eta^a$, defined implicitly 
by $g_{ab}\epsilon^a{}_A \eta^b = 0$ and $g_{ab} \eta^a \eta^b 
= -1$, up to a sign. Furthermore, note that $g_{ab} \epsilon^a{}_A \eta^b = 
(e_a  \cdot e_b) \epsilon^a{}_A \eta^b = 
 \epsilon_A \cdot (\eta^b e_b ) = 0$ so 
that $\eta^\mu = \eta^a e^\mu{}_a$. In both descriptions, 
$h^{AB}$ and $h$ denotes the inverse metric and the determinant 
of $h_{AB}$, respectively. We also define  ${\mathcal D}_A$ as 
the torsion-less  covariant derivative compatible with $h_{AB}$, see \ref{app0}.

For our purposes, it is convenient to introduce the 
following projections of the extrinsic curvature of $m$,
\beq
L_{AB}^i &=& \epsilon^a{}_A \epsilon^b{}_B K_{ab}^i
= - n^i \cdot \mathcal{D}_A \epsilon_B,
\\
L_A{}^i &=& \epsilon^a{}_A \eta^b K_{ab}^i =
- n^i \cdot \mathcal{D}_A \eta,
\eeq
in addition to 
\be 
k_{AB} = - g_{ab}\eta^a \mathcal{D}_A \epsilon^b{}_B
= k_{BA},
\ee
that is  the $\Sigma_t$ extrinsic curvature associated with the 
embedding of $\Sigma_t$ in $m$ given by $x^a = \chi^a (u^A)$.

\smallskip
\noindent
In a similar manner, in this geometrical framework the velocity vector, 
$\dot{X}^a = \partial_t X^\mu$, is tangent to the worldvolume $m$. In terms of 
the basis $\{ \epsilon^a{}_A, \eta^a\}$ the velocity  can  be written as
\be 
\dot{X}^a = N\,\eta^a + N^A \epsilon^a{}_A, 
\label{dotX}
\ee
where,  using familiar ADM terminology, 
$N$ and $N^A$ are the \textit{lapse} and the 
\textit{shift vector}, respectively. Since the lapse and the shift
vector are expressed in terms of the derivatives of $X^\mu$, {\it  i.e.}
$N = - g_{ab} \dot{X}^a \eta^b$ and $N^A = g_{ab}h^{AB} \dot{X}^a 
\epsilon^b{}_B$, neither $N$ nor $N^A$ is a canonical field variable.
Indeed, contrary to what happens in the ADM treatment for the 
metric-GR, in the treatment adopted for extended objects both 
the lapse function and the shift vector are functions of the 
phase space, and not Lagrange multipliers.

When considering the evolution of $\Sigma_t$ it is convenient to
choose first the coordinate basis $\{ \epsilon^a{}_A, \dot{X}^a \}$.
In this sense, the projections of the worldvolume metric $g_{ab}$
with respect to this basis provide immediately its ADM form. We have
\be
\begin{aligned}
g_{00} &= g_{ab} \dot{X}^a \dot{X}^b = - N^2 + N^A N^B h_{AB},
\\
g_{0A} & = g_{ab} \dot{X}^a \epsilon^b{}_A = N^B h_{AB},
\\
g_{AB} & = g_{ab} \epsilon^a{}_A  \epsilon^b{}_B = h_{AB}.
\end{aligned}
\ee
In matrix form, the induced metric and its inverse are given by
\be 
\label{gADM}
(g_{ab}) = 
\begin{pmatrix}
-N^2 + N^A N^B h_{AB} & N^A h_{AB}
\\
N^B h_{AB} & h_{AB}
\end{pmatrix},
\ee
and
\be
\label{giADM}
(g^{ab}) = \frac{1}{N^2}
\begin{pmatrix}
-1 & N^A
\\
N^A & N^2 h_{AB} - N^A N^B
\end{pmatrix},
\ee
respectively. Its determinant is given by $g = - N^2 h$.

\sk
We can make a similar ADM decomposition of the extrinsic 
curvature
\be
\begin{aligned}
K_{00}^i & = K_{ab}^i \dot{X}^a \dot{X}^b = - n^i \cdot
\ddot{X},
\\
K_{0A}^i & = K_{ab}^i \dot{X}^a \epsilon^b{}_A = - n^{i} 
\cdot \mathcal{D}_A \dot{X},
\\
K_{AB}^i &=  K_{ab}^i \epsilon^a{}_A \epsilon^b{}_B = 
- n^{i} \cdot \mathcal{D}_A \mathcal{D}_B X = L_{AB}^i.
\end{aligned}
\ee
In matrix form we have
\be 
\label{ki}
K_{ab}^i = -
\begin{pmatrix}
 n^{i} \cdot \ddot{X} &  n^i \cdot \mathcal{D}_A \dot{X}
\\
 n^i \cdot \mathcal{D}_A \dot{X} &  - L_{AB}^i 
\end{pmatrix}.
\ee

\smallskip
\noindent
The mean extrinsic curvature, $K^i = g^{ab}K_{ab}^i$, 
using (\ref{giADM}) and (\ref{ki}), becomes
\beq
K^i &=& \frac{1}{N^2} \left[ (n^i \cdot \ddot{X} )  - 2 N^A
(n^i \cdot \mathcal{D}_A \dot{X}) + (N^2h^{AB}
- N^A N^B) L_{AB}^i
\right].
\label{KADM}
\eeq
We observe the linear dependence of $K^i$ on the accelerations
of the extended object in the first term. In passing, we note that for pure 
normal evolution, $N^A = 0$, the previous expression specializes 
to
\be 
N^2 K^i = n^i \cdot \ddot{X} + N^2 L^i,
\ee
where $L^i := h^{AB} L_{AB}^i = - n^i \cdot \D^A \D_A X$,
that emphasizes the linear dependance on the acceleration.

\smallskip
\noindent
By considering the contracted integrability conditions 
associated to the Gauss-Weingarten equations (\ref{GW1}), 
the worldvolume Ricci scalar can be expressed as a sum 
of a first-order function and a divergence
term
\be
\mathcal{R} = R + k_{AB}k^{AB} - k^2 + 2 \nabla_a (k\eta^a
- \eta^b \nabla_b \eta^a),
\label{R1}
\ee
where $k=h^{AB} k_{AB}$ and $R$ is the Ricci scalar defined 
on $\Sigma$. The presence of the last term, a total divergence, 
should not come as a surprise. In metric-GR  it is the well known 
Gibbons-Hawking-York boundary term \cite{GHY1,GHY2,GHY3}, that can either be 
subtracted from the outset, or kept as in the proof of the positivity of
energy theorems by Schoen and Yau \cite{shoen1979,shoen1981}.

Alternatively,  using the integrability conditions associated
to the Gauss-Weingarten  equations (\ref{GW2}),  the induced scale curvature can be expressed as a single second-order function, when the boundary term is kept,
\be
\mathcal{R} = 2 L_i K^i - G^{ABCD} \,\Pi_{\mu\nu} \,
\mathcal{D}_A \mathcal{D}_B X^\mu \,
\mathcal{D}_C \mathcal{D}_D X^\nu 
- 2 h^{AB} \delta_{ij} 
\widetilde{\mathcal{D}}_A n^i \cdot \widetilde{\mathcal{D}}_B n^j. 
\label{R2}
\ee
The linear dependance on the acceleration is hidden through $K^i$ in the first term, see (\ref{KADM}).
Here, $\Dt_A$ denotes the covariant derivative associated 
with the connection $\sigma_A{}^{ij} := \epsilon^a{}_A 
\omega_a^{ij}$, that takes into account the rotation freedom 
of the normal vector fields, see \ref{app0}. 
Furthermore, $\Pi_{\mu\nu} := n_\mu{}^i n_{\nu\,i}$ is a symmetric normal
 projector satisfying $\Pi^\mu{}_\alpha \Pi^\alpha{}_\nu 
= \Pi^\mu{}_\nu$, and
\be
G^{ABCD} := h^{AB} h^{CD} - \frac{1}{2} (h^{AC}h^{BD} + h^{AD} h^{BC}),
\label{GABCD}
\ee
is a Wheeler-DeWitt like metric associated to $h_{AB}$. In
passing, we must bear in mind the function dependance of the normals, $n^{\mu\,i}
= n^{\mu\,i}( X^\alpha, \dot{X}^\alpha)$. Clearly, the 
canonical formulation based on the expression (\ref{R1}) 
that neglects  the divergence term, assuming a brane without boundaries, as is commonly 
done, is a different starting point  than the expression (\ref{R2}) that includes it. This second avenue is
the one taken in this paper.

\bigskip
\noindent
Since our interest lies in performing a canonical description of 
the RT model respecting its second-order nature, we 
perform the ADM decomposition of all the terms of the action (\ref{action0}) 
as follows
\be
S_{\text{\tiny RT}}[X^\mu] = \int_\mathbb{R}dt\,
L_{\text{\tiny RT}} (X^\mu{}_A, \dot{X}^\mu, \dot{X}^\mu{}_A,
\ddot{X}^\mu),
\label{action1}
\ee
where we recall that $\dot{X}^\mu$ belongs to the configuration
space from the Ostrogradsky-Hamilton viewpoint, and
\be
L_{\text{\tiny RT}} = \int_\Sigma d^p u \,
\mathcal{L}_{\text{\tiny RT}} = \int_\Sigma 
\mathcal{L}_{\text{\tiny RT}}.
\label{lag1}
\ee
For convenience in the notation, hereafter, the 
differential $d^p u$ whenever a $\Sigma$ integration 
is performed will be absorbed. 

The Lagrangian density is
\be
\mathcal{L}_{\text{\tiny RT}} = \frac{1}{2}N\sqrt{h}
\left[2 L_i K^i - G^{ABCD} \,\Pi_{\mu\nu} \,
\mathcal{D}_A \mathcal{D}_B X^\mu \,
\mathcal{D}_C \mathcal{D}_D X^\nu 
-   2 h^{AB} \delta_{ij} \widetilde{\mathcal{D}}_A 
n^i \cdot \widetilde{\mathcal{D}}_B n^j \right].
\label{lag2}
\ee
The structure of this Lagrangian  density deserves a few comments.
In the first term, through the mean extrinsic curvature, the linear acceleration dependance is hidden, see the first term in (\ref{KADM}).
The second term involves both the Wheeler-De Witt  like superspace metric and the normal projector $\Pi^{\mu\nu} = n^\mu{}^i
n^\nu_i$, defined earlier.
Finally the  last term is precisely the expression of a nonlinear sigma 
model built from $n^{\mu\,i}$, with $O ( N - p - 1)$ symmetry that 
reflects the invariance under rotations of the normal vectors
$n^{\mu}{}_i = n^\mu{}_i (X^\alpha, \dot{X}^\alpha)$, 
constrained to satisfy $n^i \cdot n^j = \delta^{ij}$. 
The Lagrangian density (\ref{lag2}) is our starting
point for obtaining the Hamiltonian formulation of GBG.

Regarding the ADM decomposition of the linear momentum density 
(\ref{pi}), using the tangential projector from $m$ onto the hypersurface 
$\Sigma$ defined as $\mathcal{H}^{ab} = h^{AB}\epsilon^a{}_A 
\epsilon^b{}_b = g^{ab} + \eta^a \eta^b$, we have
\be
\pi_\mu = - \sqrt{h} \eta^a G_{ab}g^{bc} \,e_{\mu\,c}
=  \sqrt{h} \left[ (\eta^a G_{ab} \eta^b)\eta_\mu 
- (\eta^a G_{ab} \epsilon^{b\,B})\epsilon_{\mu B}\right],
\ee
where we have considered the fact that $\sqrt{-g} = N \sqrt{h}$.
Moreover, taking into account the integrability conditions 
associated to (\ref{GW2}), the projections of the worldvolume 
Einstein tensor are given by
\beq
\eta^a G_{ab} \eta^b &=& \frac{1}{2} \left( R - k_{AB}k^{AB}
+ k^2\right),
\\
\eta^a G_{ab} \epsilon^{b\,B} &=& \D_A (k^{AB} - h^{AB}k) 
= - (L^{AB}_i - h^{AB}L_i)L_A{}^i,
\\
\epsilon^a{}_A G_{ab} \epsilon^b{}_B &=& K_i L_{AB}^i
- L_{AC}^i L^C{}_{B\,i} + L_A^i L_{B\,i} - \frac{1}{2}
\mathcal{R} \,h_{AB},
\eeq
where we recall that $R$ denotes the Ricci scalar of the hypersurface $\Sigma_t$.

%%%%%%%%%%%%%%%%%%%%%%%%%%%%%%%%%%%%%%%%%%%%%%%%%%%%%%%%%%%
\section{Ostrogradsky-Hamilton approach}
\label{sec4}
%%%%%%%%%%%%%%%%%%%%%%%%%%%%%%%%%%%%%%%%%%%%%%%%%%%%%%%%%%%

According to the Ostrogradsky-Hamilton formulation, we have 
a $4N$-dimensional phase space spanned by two conjugate pairs 
$\{ X^\mu, p_\mu; \dot{X}^\mu, P_\mu \}$ where the momenta
$p_\mu$ and $P_\mu$, conjugate to $X^\mu$ and $\dot{X}^\mu$
respectively, are defined in terms of the $\Sigma$ basis as
\beq
P_\mu &=& \frac{\delta L_{\text{\tiny RT}}}{\delta \ddot{X}^\mu}
=   \frac{\sqrt{h}}{N}  L^i  n_{\mu\,i},
\label{Pmu}
\\
p_\mu &=& \frac{\delta L_{\text{\tiny RT}}}{\delta \dot{X}^\mu}
- \partial_t P_\mu
= \pi_\mu + \partial_A \left( N^A P_\mu + \sqrt{h}
h^{AB} L_B{}^i \, n_{\mu\,i} \right),
\label{pmu}
\eeq
where $\pi_\mu$ is given by  (\ref{pi}). Note that the momenta
$P_\mu$ and $p_\mu$ are spatial densities of weight one because
the presence of the factor $\sqrt{h}$. Also, the integral over a spatial closed geometry of the momenta 
$p_\mu$ and $\pi_\mu$  differs by a boundary term. In
this sense, whereas the momenta $P_\mu$ are explicitly normal to the 
worldvolume,   the momenta $p_\mu$
are tangential to the worldvolume, up to a spatial divergence. In our analysis, we will keep 
this setting as general as possible, so that we are not restricting
our attention to a closed geometry, allowing for arbitrary boundary conditions. 

\smallskip
\noindent
In this extended phase space, the appropriate Legendre transformation is given by
$\mathcal{H}_0 := p \cdot \dot{X} + P \cdot \ddot{X} - 
\mathcal{L}_{\text{\tiny RT}}$, and it provides the canonical Hamiltonian density of weight one
\beq
\mathcal{H}_0 &=& p \cdot \dot{X} + 2N^A (P \cdot \mathcal{D}_A \dot{X}) 
+ (N^2 h^{AB} - N^A N^B) (P \cdot \mathcal{D}_A \mathcal{D}_B X)
\nonumber
\\
&+& \frac{1}{2} N \sqrt{h}\, G^{ABCD}\,
\Pi_{\mu\nu}\,\mathcal{D}_A \mathcal{D}_B X^\mu 
\mathcal{D}_C \mathcal{D}_D X^\nu
+  N \sqrt{h} h^{AB} \delta_{ij} \widetilde{\mathcal{D}}_A 
n^i \cdot \widetilde{\mathcal{D}}_B n^j,
\label{H0}
\eeq
so that the canonical Hamiltonian reads
\be
H_0[X^\mu,p_\mu; \dot{X}^\mu, P_\mu]
= \int_\Sigma \mathcal{H}_0 ( X^\mu,p_\mu, \dot{X}^\mu, P_\mu).
\label{H0c}
\ee
Note the linear dependence of the canonical 
Hamiltonian on the momenta $p_\mu$ and $P_\mu$. Classically, the physical 
momenta $p_\mu$ can take both negative and positive values  in phase space 
making the canonical Hamiltonian unbounded from below. In other words, the 
well known Ostrogradsky instabilities may be present in the dynamics of the theory (see {\it e.g.} \cite{woodard2015}). 
Also observe the absence of a quadratic term, $P^2$, that would be a signature of
an authentic second-order derivative brane model~\cite{nesterenko1988,ham-branes2004,rojas2021b}.
Moreover, $\mathcal{H}_0$ involves a highly nonlinear dependence on the configuration
variables $X^\mu$ and $\dot{X}^\mu$ through the lapse and shift 
functions as well as the last two terms in~(\ref{H0}).

\smallskip
\noindent
The presence of local symmetries manifests through the presence of 
constraints on the phase space variables. In principle, we can determine them by computing 
first the null eigenvectors of the Hessian matrix. In this case,
this vanishes identically,
\be
H_{\mu\nu} = \frac{\delta^2 L_{\text{\tiny RT}}}{\delta \ddot{X}^\mu 
\delta \ddot{X}^\nu} = 0.
\ee
This feature characterize theories affine in the acceleration \cite{CMR2016,rojas2021b}. 
The rank of the Hessian matrix is zero which is a signal 
that the phase space is fully constrained, {\it i.e.} we have the
 presence of $N$ primary constraints. Clearly, we cannot
invert any of the accelerations $\ddot{X}^\mu$ in favour of the phase
space variables so that the definition itself of the momenta $P_\mu$ (\ref{Pmu})
provides the set of $N$ primary constraints densities
\be
\mathcal{C}_\mu := P_\mu -  \frac{\sqrt{h}}{N}
L^i
n_{\mu\,i} = 0.
\ee
A more manageable approach to the computations with these 
constraints, without affecting their content, is to exploit the 
intrinsic geometric nature of the system. Indeed, using the 
tangential projector from $\mathcal{M}$ onto the hypersurface 
$\Sigma$, $\mathcal{H}^{\mu\nu} = h^{AB} \epsilon^\mu{}_A 
\epsilon^\nu{}_B = \eta^{\mu\nu} + \eta^\mu \eta^\nu - n^{\mu\,i}
n^\nu{}_i$, written in terms of the hypersurface $\Sigma$ basis 
$\{ \dot{X}^\mu, \epsilon^\mu{}_A, n^\mu{}_i \}$ we can rewrite 
them as $\mathcal{C}_\mu = \eta_{\mu\nu} \mathcal{C}^\nu
= \mathcal{C}_1\,\dot{X}_\mu + \mathcal{C}_A\,
\epsilon_\mu{}^A + \mathcal{C}_i\,n_{\mu}{}^i = 0$,
where we have $\eta^\mu = e^\mu{}_a \eta^a = (\dot{X}^\mu 
- N^A \epsilon^\mu{}_A)/N$. This linear combination helps to 
identify a set of equivalent primary constraints densities
\beq
\mathcal{C}_1 &:=& P \cdot \dot{X} = 0,
\label{C1}
\\
\mathcal{C}_A &:=& P \cdot \partial_A X = 0,
\label{CA}
\\
\mathcal{C}_i &:=& P \cdot n_i -
  \frac{\sqrt{h}}{N} L_i = 0.
\label{Ci}
\eeq
We will see below that these constraints do generate  the 
expected  local gauge transformations.

\smallskip
\noindent
It is convenient to turn these constraints densities into 
functions in the phase space $\Gamma$. To do this, we smear out the constraints (\ref{C1}), 
(\ref{CA}) and (\ref{Ci}) by test fields $\lambda$, $\lambda^A$ 
and $\phi^i$ defined on $\Sigma_t$, 
and then we integrate them over the entire spatial hypersurface
$\Sigma$ with
\beq
\mathcal{S}_\lambda &:=& \int_\Sigma \lambda \,P \cdot 
\dot{X},
\label{C1f}
\\
\mathcal{V}_{\vec{\lambda}} &:=& \int_\Sigma\lambda^A P 
\cdot \partial_A X,
\label{CAf}
\\
\mathcal{W}_{\vec{\phi}} &:=&
\int_\Sigma\phi^i \left[ P \cdot n_i + 
\frac{\sqrt{h}}{N} L_i
\right].
\label{Cif}
\eeq

\smallskip
\noindent
According to the Dirac-Bergmann procedure for constrained systems, 
the time evolution in the phase space $\Gamma$ is generated by the total Hamiltonian
\be
H [X^\mu,p_\mu; \dot{X}^\mu,P_\mu] = H_0 + \mathcal{S}_\lambda + 
\mathcal{V}_{\vec{\lambda}} + \mathcal{W}_{\vec{\phi}}.
\label{Ham}
\ee

\smallskip
\noindent
The time evolution of any phase space function $F$ 
is given by
\be
\partial_t F = \dot{F} \approx \{ F, H \},
\label{Fdot}
\ee
where we have used the Ostrogradsky-Poisson bracket [PB] appropriate
for second-order derivative theories
\be
\{ F , G \} = \int_\Sigma
\left[ 
\frac{\delta F}{\delta X} \cdot \frac{\delta G}{\delta p} + 
\frac{\delta F}{\delta \dot{X}} \cdot \frac{\delta G} {\delta P}
- ( F \longleftrightarrow G)
\right],
\label{PB}
\ee
with $F, G \in \Gamma$ as well as using the \textit{weak} 
equality symbol $\approx$ that means that after evaluating the PB, then 
one  imposes the vanishing of the constraints, $\mathcal{C}_{1,A,i}= 0$ \cite{HT1992}. Hence, we can 
use (\ref{Fdot}) to efficiently compute the time evolution 
of any constraint function. Before we do this, we note that 
under the PB structure the primary constraints functions 
(\ref{C1f}-\ref{Cif}), are in involution with each other. 
We have
\be
\begin{array}{ll}
\{ \mathcal{S}_\lambda , \mathcal{S}_{\lambda'} \} = 0,
& \qquad \qquad \qquad
\{ \mathcal{V}_{\vec{\lambda}}, \mathcal{V}_{\vec{\lambda}'} \} = 0,
\\
\{ \mathcal{S}_\lambda, \mathcal{V}_{\vec{\lambda}} \}
= \mathcal{V}_{\vec{\lambda}'}, \qquad \,\,\,\,\lambda^{'A} = \lambda \lambda^A,
& \qquad \qquad \qquad
\{ \mathcal{V}_{\vec{\lambda}}, \mathcal{W}_{\vec{\phi}} \}
= 0,
\\
\{ \mathcal{S}_\lambda , \mathcal{W}_{\vec{\phi}} \} = 
\mathcal{W}_{\vec{\phi}^{\,'}}, \qquad
\phi^{'i} = \lambda \phi^i,
& \qquad \qquad \qquad 
\{ \mathcal{W}_{\vec{\phi}}, \mathcal{W}_{\vec{\phi}^{\,'}} \}
= 0,
\end{array}
\label{algebra1}
\ee
where the  functional derivatives of the primary constraints 
are
\begin{equation}
\begin{array}{ll}
\frac{\delta \mathcal{S}_\lambda}{\delta \dot{X}^\mu}
= \lambda\,P_\mu ,
&
\frac{\delta \mathcal{S}_\lambda}{\delta P_\mu}
= \lambda\,\dot{X}^\mu,
\\
\frac{\delta \mathcal{V}_{\vec{\lambda}}}{\delta X^\mu}
= - \partial_A (\lambda^A P_\mu)
= - \pounds_{\vec{\lambda}}P_\mu,
&
\frac{\delta \mathcal{V}_{\vec{\lambda}}}{\delta P_\mu}
= \lambda^A \partial_A X^\mu
=  \pounds_{\vec{\lambda}}X^\mu,
\\
\frac{\delta \mathcal{W}_{\vec{\phi}}}{\delta \dot{X}^\mu}
= -  \phi^i \frac{\sqrt{h}}{N^2}L_i\,\eta_\mu
-  \phi^i \frac{\sqrt{h}}{N^2} k\,n_{\mu\,i},
&
\frac{\delta \mathcal{W}_{\vec{\phi}}}{\delta P_\mu}
= \phi^i\,n^\mu{}_i,
\\
\frac{\delta \mathcal{W}_{\vec{\phi}}}{\delta X^\mu}
= 
- \mathcal{D}_A \left(  \phi^i \frac{\sqrt{h}}{N^2}
N^AL_i\,\eta_\mu \right)
+ \mathcal{D}_A \left(  \phi^i \frac{\sqrt{h}}{N} 
h^{AB} L_i \,\epsilon_{\mu B}\right)
& 
\\
\,\,\,\,\,\,\qquad - \mathcal{D}_A \left( 2 \phi^i
\frac{\sqrt{h}}{N}L^{AB}_i \epsilon_{\mu B}\right)
- \mathcal{D}_A \left(  \phi^i \frac{\sqrt{h}}{N^2}
N^A k \,n_{\mu\,i} \right)
&
\\
\,\,\,\,\,\,\qquad + \mathcal{D}_A \mathcal{D}_B \left( 
 \phi^i \frac{\sqrt{h}}{N} h^{AB}n_{\mu\,i}
\right).
&
\end{array}
\end{equation}

\sk
In order for the primary constraints to be consistent, following 
the Dirac-Bergmann recipe, extended to higher-order Lagrangians, 
we require that their time evolution be vanishing. This procedure 
gives rise to the secondary constraints densities
\beq
C_1 &:=& \mathcal{H}_0 = 0,
\label{CC1}
\\
C_A &:=& p \cdot \partial_A X + P \cdot \partial_A \dot{X} = 0,
\label{CCA}
\\
C_i &:=& p \cdot n_i - n_i \cdot \partial_A \left( 
N^A P +  \sqrt{h}h^{AB}L_B{}^j\,n_j \right) = 0.
\label{CCi}
\eeq
It is worth mentioning that these constraints can also be obtained by 
projecting the momenta $p_\mu$ given by (\ref{pmu}), along the $\Sigma_t$ 
basis $\{ \dot{X}^\mu, \epsilon^\mu{}_A, n^\mu{}_i\}$.

\smallskip
\noindent
As above, we turn the local secondary  constraints into  secondary constraint functions in the phase space  $\Gamma$ 
by smearing them  by the test fields 
$\Lambda$, $\Lambda^A$ and $\Phi^i$, defined on $\Sigma_t$, and integrating over $\Sigma$,
\beq
S_\Lambda &:=& \int_\Sigma \Lambda  \mathcal{H}_0 ,
\label{CC1f}
\\
V_{\vec{\Lambda}} &:=& \int_\Sigma \Lambda^A \left(
p \cdot \partial_A X + P \cdot \partial_A \dot{X} \right),
\label{CCAf}
\\
W_{\vec{\Phi}} &:=& \int_\Sigma
\Phi^i \left[ 
p \cdot n_i - n_i \cdot \partial_A \left(
N^A P +  \sqrt{h}h^{AB}L_B{}^j\,n_j \right)
\right].
\label{CCif}
\eeq
Some of the functional derivatives of the secondary 
constraints are given by
\begin{equation}
\begin{array}{ll}
\frac{\delta V_{\vec{\Lambda}}}{\delta \dot{X}^\mu}
= - \partial_A (\Lambda^A P_\mu) = - 
\pounds_{\vec{\Lambda}}P_\mu,
& \qquad
\frac{\delta V_{\vec{\Lambda}}}{\delta P_\mu} 
= \Lambda^A \partial_A \dot{X}^\mu = \pounds_{\vec{\Lambda}}
\dot{X}^\mu,
\\
\frac{\delta V_{\vec{\Lambda}}}{\delta X^\mu}
= - \partial_A (\Lambda^A p_\mu) = - 
\pounds_{\vec{\Lambda}}p_\mu,
& \qquad
\frac{\delta V_{\vec{\Lambda}}}{\delta p_\mu} 
= \Lambda^A \partial_A X^\mu = \pounds_{\vec{\Lambda}}X^\mu,
\end{array}
\nonumber
\end{equation}
and {\footnotesize
\begin{equation}
\begin{aligned}
\frac{\delta W_{\vec{\Phi}}}{\delta p_\mu} & = \Phi^i n^\mu{}_i,
\qquad \qquad \qquad \qquad \qquad \qquad \quad
\frac{\delta W_{\vec{\Phi}}}{\delta P_\mu} = N^A \mathcal{D}_A
(\Phi^i n^\mu{}_i) = \mathcal{L}_{\vec{N}} ( \Phi^i n^\mu{}_i),
\\
\frac{\delta W_{\vec{\Phi}}}{\delta \dot{X}^\mu} & = (P \cdot n_i)
\widetilde{\mathcal{D}}_A \Phi^i h^{AB} \epsilon_{\mu B}
+ \frac{1}{N} \Dt_A \left( \Dt_B \Phi^i\, \sqrt{h} h^{AB}
\right) n_{\mu\,i}
+  \frac{\sqrt{h}}{N}\,\Phi_i  L_A{}^i L^A{}_j\,n_\mu{}^j
\\
&+
\frac{1}{N} (p\cdot \eta)\Phi_i \,n_\mu{}^i
- \frac{1}{N} \left[ N^A L_A{}^j (P \cdot n_j) 
+ \alpha \sqrt{h} \,L_A{}^j L^A{}_j \right]\Phi_i \,n_\mu{}^i,
\\
\frac{\delta W_{\vec{\Phi}}}{\delta X^\mu} &= 
\D_A \widetilde{W}^A,
\end{aligned}
\nonumber
\end{equation}}
where {\footnotesize 
\beq
\widetilde{W}^A &=&  h^{AB} (p \cdot \epsilon_B)\Phi_i n_\mu{}^i +
\frac{N^A}{N} (p \cdot \eta)\Phi_i n_\mu{}^i
+ \Dt_B \Phi_i (P \cdot n^i)h^{AB} \dot{X}_\mu 
\n
\\
&-& \left[ 
 h^{AB}N^C L_{BC}^j (P \cdot n_j) +  \sqrt{h} h^{AB} L^C{}_j L_{BC}^j
\right]\Phi_i n_\mu{}^i
+ 2 N^{(A}h^{B)C} (P \cdot n^i) \Dt_C \Phi_i\,\epsilon_{\mu\,B}
\n
\\
&-& \left[ 
\frac{N^AN^B}{N}L_B{}^j (P \cdot n_j) + \sqrt{h}\frac{N^A}{N}
L^B{}_j L_B{}^j \right]\Phi_i n_{\mu}{}^i
+
2 \sqrt{h} h^{B(C} L^{A)}{}_i \Dt_B \Phi^i\,\epsilon_{\mu\,C}
\n
\\
&-& \sqrt{h} h^{AB} L^C{}_i \Dt_C \Phi^i \,\epsilon_{\mu\,B}
+ \sqrt{h} h^{AB}h^{CD} L_{BC}^i \Dt_D \Phi_i \,\eta_\mu 
- \sqrt{h} k^{AB} \Dt_B \Phi_i\,n_\mu{}^i
\n
\\
&+& \sqrt{h}\frac{N^A}{N}h^{BC} \Dt_B \Dt_C \Phi_i\,n_\mu{}^i
+ \sqrt{h} h^{AB} L^C{}_j L^i_{BC} \Phi_i \,n_\mu{}^j
+ \sqrt{h}\frac{N^A}{N} L^B{}_j L_B{}^i \Phi_i\,n_\mu{}^j.
\eeq}
is a vector field with density weight $+1$. The functional 
derivatives of $S_\Lambda$ have been listed in \ref{app1}. The 
following step is to check that the evolution 
in time of the secondary constraints does not 
generate further, tertiary constraints.  
By using the PB algebra listed in (\ref{algebra1}) 
we do not find tertiary constraints so that the Dirac-Bergmann 
algorithm terminates.

\smallskip
\noindent
We make the following remarks. The constraints (\ref{C1}) and (\ref{CA})
are characteristic of second-order derivative brane models and just 
involve the momenta $P_\mu$. Geometrically, they may be interpreted
as a consequence of the orthonormality of the worldvolume basis.
On the contrary, the constraints (\ref{CC1}) and (\ref{CCA}) involve
all the phase space variables. Regarding these, (\ref{CC1})
reflects the vanishing of the canonical Hamiltonian, that is expected because of the invariance under worldvolume  reparametrization of the theory. 
Indeed, it generates diffeomorphisms out of $\Sigma$ onto the worldvolume. 
On the other hand, (\ref{CCA}) generates diffeomorphisms tangential 
to $\Sigma$. This fact can be verified by considering the PB with the phase space variables as we will see shortly.
These two constraints should be recognizable to the reader familiar with the ADM formulation of General Relativity \cite{wald}.
Regarding the remaining constraints (\ref{Ci}) and (\ref{CCi}), 
the first one represents a way of expressing the trace of the 
spatial-spatial projection of the extrinsic curvature, $L^i$, in terms 
of the phase space variables and the second one reflects the 
orthogonality between the physical momenta $\pi_\mu$ and the normal 
vectors to the worldvolume. In other words, the constraints (\ref{Ci}) and 
(\ref{CCi}) are characteristic of brane models linear in accelerations.

\smallskip
\noindent
The remaining PB between the primary and secondary 
constraints, as well as between the secondary constraints themselves, 
are listed in \ref{app2}. These relationships are helpful in identifying 
the first and second class constraints of the model.

%%%%%%%%%%%%%%%%%%%%%%%%%%%%%%%%%%%%%%%%%%%%%%%%%%%%%%
\subsection{Hamilton's equations}
%%%%%%%%%%%%%%%%%%%%%%%%%%%%%%%%%%%%%%%%%%%%%%%%%%%%

Here we obtain the field equations in the Hamiltonian 
formulation. This computation is helpful in order to fix 
some Lagrange multipliers that appear as test functions in the definition of the constraints as functions in phase space  in terms of the phase space variables. In addition it provides a check, as it reproduces the form of the momenta $P_\mu, p_\mu$ given by (\ref{Pmu}) and (\ref{pmu}),
respectively.

By considering the functional derivatives listed
before and in \ref{app1}
as well as the Hamiltonian (\ref{Ham}) we have first that
\be
\partial_t X^\mu  = \{ X^\mu , H \} = \frac{\delta H_0}{\delta p_\mu}
= \dot{X}^\mu\,.
\ee
This result is obvious since the only dependence on $p_\mu$
is through the term $p\cdot \dot{X}$ appearing in $H_0$.
Secondly, we 
compute
\beq
\partial_t \dot{X}^\mu &=& \{ \dot{X}^\mu , H \} =
\frac{\delta H_0}{\delta P_\mu} + \frac{\delta \mathcal{S}_\lambda}{\delta P_\mu} 
+ \frac{\delta \mathcal{V}_{\vec{\lambda}}}{\delta P_\mu}
+ \frac{\delta \mathcal{W}_{\vec{\phi}}}{\delta P_\mu},
\nonumber 
\\
&=& 2N^A \D_A \dot{X}^\mu + (N^2 h^{AB} - N^A N^B)
\D_A \D_B X^\mu + \lambda \dot{X}^\mu + \lambda^A 
\epsilon^\mu{}_A + \phi^i n^\mu{}_i.\,\,\,\,
\label{id1}
\eeq
By contracting (\ref{id1}) with the momenta $P_\mu$ and considering
the identity (\ref{KADM}) and the primary constraint densities 
(\ref{C1}), (\ref{CA}) and (\ref{Ci}), we identify
\be
\phi^i = N^2 K^i.
\label{LMi}
\ee
In order to fix the remaining Lagrange multipliers, it is useful
to recall an important identity relating the acceleration in terms 
of the $\Sigma_t$ basis \cite{ham-branes2004},
\be
\ddot{X}^\mu = (\dot{N}_A + N \D_A N - N^B\D_A N_B) 
\epsilon^{\mu A} + (\dot{N} + N^A\D_A N + N^A N^B k_{AB})\eta^\mu
+ (n^i \cdot \ddot{X}) n^\mu{}_i.
\label{id2}
\ee
As before, by considering (\ref{id2}) and the primary constraints, 
when contracting (\ref{id1}) with  $\eta$ and $\epsilon_A$
yields
\beq
\lambda &=& \D_A N^A - \frac{N^2}{\sqrt{h}} \eta^a \nabla_a \left( 
\frac{\sqrt{h}}{N}\right) =  \frac{1}{N} \left( \dot{N} - 
N^A \D_A N - N^2 k \right),
\label{LM1}
\\
\lambda^A &=& N\D^A N - N^A \D_B N^B + \frac{N^2}{\sqrt{h}} \eta^a \nabla_a \left( 
\frac{\sqrt{h}N^A}{N}\right),
\label{LMA}
\eeq
where we have used the time derivative of the spatial metric,
$\dot{h}_{AB} = 2Nk_{AB} + 2 \D_{(A} N_{B)}$ and its determinant,
$\partial_t(\sqrt{h}) = \sqrt{h} (NK + \D_A N^A)$.
It is worthwhile to mention that the Lagrange multipliers (\ref{LM1})
and (\ref{LMA}) are inherent to second-order derivative brane 
models \cite{ham-branes2004,rojas2021}.

\smallskip
\noindent
We turn now to compute the time evolution of the momenta $P_\mu$. We obtain the lengthy
expression
\beq
\partial_t P_\mu &=& \{ P_\mu , H \} = -\frac{\delta H_0}{\delta 
\dot{X}^\mu} - \frac{\delta \mathcal{S}_\lambda}{\delta \dot{X}^\mu}
- \frac{\delta \mathcal{W}_{\vec{\phi}}}{\delta \dot{X}^\mu},
\nonumber
\\
&=& - p_\mu 
- 2 (P \cdot \D_A \dot{X})h^{AB} \epsilon_{\mu\,B} 
+ \D_A (2 N^A\,P_\mu) 
+ 2 N h^{AB} (P \cdot \D_A \D_B X)\, \eta_\mu
\nonumber 
\\
&+&
2 N^B (P \cdot \D_A \D_B X) h^{AC}\epsilon_{\mu C}
+ \frac{1}{2} \sqrt{h} G^{ABCD} \Pi_{\alpha\beta}
\mathcal{D}_A \mathcal{D}_B X^\alpha \mathcal{D}_C \mathcal{D}_D
X^\beta\,\eta_\mu
\nonumber 
\\
&-&   \sqrt{h}\,G^{ABCD}L_{AB}^i k_{CD}\,n_{\mu\,i}
+   \sqrt{h} h^{AB} \delta_{ij}
\left( \Dt_A n^i \cdot \Dt_B n^j \right)\eta_\mu
\nonumber 
\\
&+& \left[ 
\Dt_A \left( 2  N \sqrt{h} h^{AB} \delta_{ij}
\Dt_B n^j \right) \cdot \eta \right]\frac{1}{N} n_\mu{}^i
-\lambda P_\mu +  \phi^i \frac{\sqrt{h}}{N^2} L_i \,\eta_\mu 
+  \phi^i \frac{\sqrt{h}}{N^2} k\,n_{\mu\,i}
\nonumber 
\eeq
By inserting (\ref{LMi}) and (\ref{LM1}) into 
the previous expression
we get 
\beq
p_\mu &=&  \left \lbrace \alpha \sqrt{h} \left[  L_i K^i  - 2 L_i L^i 
+  \frac{1}{2}  G^{ABCD} \Pi_{\alpha\beta}
\D_A \D_B X^\alpha \D_C \D_D X^\beta
\right. \right.
\nonumber
\\
&+& \left. \left.  h^{AB} \delta_{ij}
\left( \Dt_A n^i \cdot \Dt_B n^j \right) \right]\eta_\mu 
\right. 
\nonumber 
\\
&+& \left.  \sqrt{h} \left[ k K^i + \Dt_A \left( 2  N h^{AB} 
\delta_{ij} \Dt_B n^j \right) \cdot \left(\frac{1}{N} \eta \right)
- G^{ABCD} L_{AB}^i k_{CD} \right] n_{\mu\,i}
\right.
\nonumber
\\
&+& \left. 2 \sqrt{h} h^{AB}L_i L_A{}^i\,\epsilon_{\mu\,B}
- \frac{1}{N} (\dot{N} - N^A \D_A N - N^2 k)P_\mu
+ \D_A (2N^A P_\mu) \right\rbrace 
\nonumber
\\
&-& \partial_t P_\mu.
\eeq
This expression matches the definition of $p_\mu$  (\ref{pmu}) for a higher derivative theory once we identify the term 
inside the curly brackets on the r.h.s. as $\delta L_{\text{\tiny RT}}/\delta \dot{X}^\mu$.
Also, this expression exhibits the linear dependence of the momenta
$p_\mu$ on the accelerations of the extended object \cite{rojas2021}.
So far, the Hamilton's equations do their job in that they correctly reproduce the expressions for the momenta as well the expressions for the velocity and the accelerations of the extended object. 
Finally, the time evolution of momenta $p_\mu$ 
\be
\partial_t p_\mu = \{ p_\mu, H\} = - \frac{\delta H_0}{\delta X^\mu}
- \frac{\delta \mathcal{S}_\lambda}{\delta X^\mu} - 
\frac{\delta \mathcal{V}_{\vec{\lambda}}}{\delta X^\mu} 
- \frac{\delta \mathcal{W}_{\vec{\phi}}}{\delta X^\mu},
\ee
are nothing but the field equations of the model (\ref{eom0})
in its canonical form as it may shown after a very long, but 
straightforward, calculation after introducing the explicit form of the Lagrange multipliers 
(\ref{LMi}), (\ref{LM1}) and (\ref{LMA}). We refrain from writing it down explicitly because it is too cumbersome.

%%%%%%%%%%%%%%%%%%%%%%%%%%%%%%%%%%%%%%%%%%%%%%%%%%%%%%%
\section{First- and second-class constraints}
\label{sec5}
%%%%%%%%%%%%%%%%%%%%%%%%%%%%%%%%%%%%%%%%%%%%%%%%%%%%%%%

In order to characterize the constraint 
surface we need to separate 
both the primary and secondary constraints into first and second class
constraints. To begin with, we relabel the constraints functions
as follows
\be
\varphi_I := \{ \mathcal{W}_{\vec{\phi}}, \mathcal{S}_\lambda, 
\mathcal{V}_{\vec{\lambda}}, S_\Lambda, V_{\vec{\Lambda}},
W_{\vec{\Phi}} \}, \qquad \qquad \quad I = 1,2, \ldots, 6.,
\ee
where we have chosen a convenient order for them.
Then, we turn to construct the antisymmetric matrix composed 
of the PB of all the constraint functions, $\Omega_{IJ} := 
\{ \varphi_I, \varphi_J \}$. Explicitly, the matrix
$\Omega_{IJ}$ reads, weakly on the constraint surface,
\be
(\Omega_{IJ}) \approx
\begin{pmatrix}
0 & 0 & 0 & 0 & 0 & \mathcal{C}
\\
0 & 0 & 0 & 0 & 0 & \mathcal{A}
\\
0 & 0 & 0 & 0 & 0 & \mathcal{B}
\\
0 & 0 & 0 & 0 & 0 & \mathcal{D}
\\
0 & 0 & 0 & 0 & 0 & \mathcal{E}
\\
-\mathcal{C} & -\mathcal{A} & -\mathcal{B} 
& -\mathcal{D} & - \mathcal{E} & \mathcal{F}
\end{pmatrix},
\label{Omega}
\ee
where the nonvanishing entries $\mathcal{A}, \mathcal{B},
\mathcal{C}, \mathcal{D}, \mathcal{E}$ and $\mathcal{F}$
are defined in \ref{app2}. The rank of this matrix
is 2, thus pointing out the existence of two second-class
constraint functions. To select these it is necessary to 
determine first the 4 zero modes $\omega^I_{(u)}$ with 
$z = 1,2,3,4$, so that $\Omega_{IJ} \omega^J_{(u)} = 0$. 
These can be taken as follows
\be
\omega^I_{(1)} =
\begin{pmatrix}
- \mathcal{A}/\mathcal{C}
\\
1
\\
0
\\
0
\\
0
\\
0
\end{pmatrix},
\quad
\omega^I_{(2)} =
\begin{pmatrix}
- \mathcal{B}/\mathcal{C}
\\
0
\\
1
\\
0
\\
0
\\
0
\end{pmatrix},
\quad
\omega^I_{(3)} =
\begin{pmatrix}
- \mathcal{D}/\mathcal{C}
\\
0
\\
0
\\
1
\\
0
\\
0
\end{pmatrix},
\quad
\omega^I_{(4)} =
\begin{pmatrix}
- \mathcal{E}/\mathcal{C}
\\
0
\\
0
\\
0
\\
1
\\
0
\end{pmatrix}.
\label{modes1}
\ee

\smallskip
\noindent
With these the functions $\mathsf{F}_u := \omega^I_{(u)} 
\varphi_I$ are first-class constraints,
\begin{equation}
\begin{array}{ll}
\mathsf{F}_1 = \mathcal{S}_\lambda 
- \frac{\mathcal{A}}{\mathcal{C}} \mathcal{W}_{{\phi}},
& \qquad \qquad \qquad
\mathsf{F}_2 = \mathcal{V}_{\vec{\lambda}} 
- \frac{\mathcal{B}}{\mathcal{C}} \mathcal{W}_{{\phi}},
\\
\mathsf{F}_3 = S_\Lambda - \frac{\mathcal{D}}{\mathcal{C}} \mathcal{W}_{{\phi}},
& \qquad \qquad \qquad
\mathsf{F}_4 = V_{\vec{\Lambda}} 
- \frac{\mathcal{E}}{\mathcal{C}} \mathcal{W}_{{\phi}}.
\end{array}
\label{FC1}
\end{equation}

\smallskip
\noindent
To formally obtain the second-class constraints we proceed 
as follows. If we choose a set of linearly independent 
vectors, $\omega^I_{(u')}$ with $u'=5,6$, such that they do not 
depend on the vectors $\omega^I_{(u)}$ and satisfy the 
condition $\det (\omega^I_{(I')}) \neq 0$ with $I' = (u,u')$, 
then the functions $\mathsf{S}_{u'} := \omega^I_{(u')} \varphi_I$ are 
second-class constraints \cite{gitman1990}. Indeed, by 
choosing
\be
\omega^I_{(5)} =
\begin{pmatrix}
1
\\
0
\\
0
\\
0
\\
0
\\
0
\end{pmatrix},
\quad \text{and} \quad
\omega^I_{(6)} =
\begin{pmatrix}
0
\\
0
\\
0
\\
0
\\
0
\\
1
\end{pmatrix},
\label{modes2}
\ee
we observe that the previously mentioned conditions are 
satisfied. Then,
\beq
\mathsf{S}_5 &=& \mathcal{W}_{\vec{\phi}},
\\
\mathsf{S}_6 &=& W_{\vec{\Phi}},
\eeq
are second-class constraints. 

\smallskip
\noindent
The constraints $\mathsf{F}_u$ and $\mathsf{S}_{u'}$ define 
an equivalent representation of the constrained phase space.
In this new framework for the constraint surface we can introduce
the matrix elements $\mathsf{S}_{u'v'} := \{ \mathsf{S}_{u'}, 
\mathsf{S}_{v'}\}$ with $u',v'=5,6$., and its inverse matrix 
components $(\mathsf{S}^{-1})^{u'v'}$, given by
\be
(\mathsf{S}_{u'v'}) = 
\begin{pmatrix}
0 & \mathcal{C}
\\
-\mathcal{C} & \mathcal{F}
\end{pmatrix},
\qquad \text{and} \qquad
\left( (\mathsf{S}^{-1})^{u'v'}\right) = \frac{1}{\mathcal{C}^2}
\begin{pmatrix}
\mathcal{F} & - \mathcal{C}
\\
\mathcal{C} & 0
\end{pmatrix},
\label{Smatrices}
\ee
respectively. According to the theory for constrained systems, 
the matrix $(\mathsf{S}^{-1})^{u'v'}$ allows us to introduce 
the Dirac bracket in the usual way
\be
\label{DB}
\{ F, G \}_D := \{F, G \} - \{ F, \mathsf{S}_{u'} \}
(\mathsf{S}^{-1})^{u'v'} \{ \mathsf{S}_{v'}, G \}.
\ee
Once we have formally determined the second-class constrictions, 
we can set them strongly equal to zero, so that these merely 
becomes identities serving to express some phase space variables 
in terms of others. Accordingly, the first-class constraints 
(\ref{FC1}) reduce to 
\be
\label{FC2}
\begin{array}{ll}
\mathsf{F}_1 = \mathcal{S}_\lambda,
& \qquad \qquad
\mathsf{F}_3 = S_\Lambda,
\\
\mathsf{F}_2 = \mathcal{V}_{\vec{\lambda}},
& \qquad \qquad
\mathsf{F}_4 = V_{\vec{\Lambda}},
\end{array}
\ee
as expected. It is worth observing that each of the first-class 
constraints functions, $\mathsf{F}_2$ and $\mathsf{F}_4$, 
includes $p$ primary constraints and $p$ secondary constraints, 
respectively.  Similarly, each of the second-class constraints 
functions, $\mathsf{S}_5$ and $\mathsf{S}_6$, includes $(N - p - 1)$ 
primary constraints and $(N - p -1)$ secondary constraints, respectively. 
In this sense, the counting of the physical degrees of freedom
[dof] is as follows: 2 dof = (total number of canonical variables)
- 2 (number of first-class constraints) - (number of second-class 
constraints). That is, dof $= N - p - 1 = i$. Hence, there are 
$i$ degrees of freedom, one for each normal vector of the worldvolume.
This number agrees with the number of physical transverse motions $\sigma^i := n^i \cdot 
\delta X$ characterizing first-order derivative brane models, as expected. 

\smallskip
\noindent
With support with the gauge transformations that generate the 
first-class constraints, it is convenient to name 
$\mathcal{S}_\lambda$ the \textit{shift constraint} while 
$\mathcal{V}_{\vec{\lambda}}$ will be referred to as the 
\textit{primary vector constraint}. In the same spirit, 
$S_\Lambda$ and $V_{\vec{\Lambda}}$ may be thought
of as being the \textit{scalar} and \textit{secondary vector}
constraint, respectively, in comparison to the ones appearing 
in a canonical analysis of the Dirac-Nambu-Goto model 
\cite{ham-branes2004}. 

%%%%%%%%%%%%%%%%%%%%%%%%%%%%%%%%%%%%%%%%%%%%%%%%%%%%%%
\subsection{Algebra of constraints}
%%%%%%%%%%%%%%%%%%%%%%%%%%%%%%%%%%%%%%%%%%%%%%%%%

Under the Dirac bracket, the algebra spanned by the 
first-class constraints is
\begin{subequations}
\begin{align}
\{ \mathcal{S}_{\lambda}, \mathcal{S}_{\lambda'} \}_D & = 0,
\label{PBa}
\\
\{ \mathcal{S}_{\lambda}, \mathcal{V}_{\vec{\lambda}} \}_D & = 
\mathcal{V}_{\vec{\lambda}_1},
\label{PBb}
\\
\{ \mathcal{S}_{\lambda}, S_{\Lambda} \}_D & = - \mathcal{S}_{\lambda_1}
- S_{\Lambda_1},
\label{PBc}
\\
\{ \mathcal{S}_{\lambda}, V_{\vec{\Lambda}} \}_D & =
- \mathcal{S}_{\pounds_{\vec{\Lambda}}\lambda},
\label{PBd}
\\
\{ \mathcal{V}_{\vec{\lambda}}, \mathcal{V}_{\vec{\lambda}^{\,'}} 
\}_D & = 0,
\label{PBe}
\\
\{ \mathcal{V}_{\vec{\lambda}}, S_\Lambda \}_D & =
\mathcal{S}_{\pounds_{\vec{\lambda}}\Lambda} - 
\mathcal{V}_{\vec{\lambda}_2} - V_{\vec{\Lambda}_1},
\label{PBf}
\\
\{ \mathcal{V}_{\vec{\lambda}}, V_{\vec{\Lambda}} \}_D & =
\mathcal{V}_{[\vec{\lambda},\vec{\Lambda}]},
\label{PBg}
\\
\{ S_\Lambda, S_{\Lambda'} \}_D & = \mathcal{S}_{\lambda_2},
\label{PBh}
\\
 \{ S_\Lambda , V_{\vec{\Lambda}} \}_D & = 
- S_{\pounds_{\vec{\Lambda}}\Lambda} + \mathcal{V}_{\vec{\lambda}_3},
\label{PBi}
\\
\{ V_{\vec{\Lambda}}, V_{\vec{\Lambda}^{\,'}} \}_D & = 
V_{[\vec{\Lambda}, \vec{\Lambda}^{\,'}]},
\label{PBj}
\end{align}
\end{subequations}
where we have introduced
\begin{equation}
\begin{array}{ll}
\lambda_1^A = \lambda \lambda^A ,
& \qquad \quad 
\lambda^A_2 = 2\Lambda N^B \D_B \lambda^A,
\\
\lambda_1 = 2\Lambda \pounds_{\vec{N}}\lambda,
& \qquad \quad
\lambda_2 = (N^2 h^{AB} - N^A N^B) (\Lambda \D_A \D_B \Lambda'
- \Lambda' \D_A \D_B \Lambda),
\\
\Lambda^A_1 = \Lambda \lambda^A,
& \qquad \quad
\lambda^A_3 = \Lambda (N^2 h^{AB} - N^A N^B)
(\D_A \D_B \Lambda^C + R_{ADB}{}^C \Lambda^D),
\\
\Lambda_1 = \lambda \Lambda.
&
\end{array}
\label{test}
\end{equation}
This algebra is  equivalent to the algebra under 
under the PB, once we apply the property
$\{ F, \mathsf{F}_u \} \approx \{ F, \mathsf{F}_u \}_D$,
for any phase space function $F$.

\smallskip
\noindent
The geometrical interpretation of this algebra can be illustrated as follows. 
Let us begin with (\ref{PBh}). We observe that two different 
orderings of the scalar constraints may only differ by a shift 
transformation which means that the time evolution with the 
scalar constraint is unique up to a rescaling. From 
(\ref{PBi}) we note that the PB of a vector with a scalar 
constraint is a scalar constraint with a test field given by 
the Lie derivative of the parameter $\Lambda$ along the vector 
field $\vec{\lambda}$; this is accompanied by tangential 
deformations provide by the primary vector constraint. Relationship
(\ref{PBj}) shows that secondary vector constraints generate 
a proper subalgebra  of their own, {\it i.e.}  it exhibits the 
invariance under reparametrizations of the theory. Regarding
(\ref{PBa}) and (\ref{PBe}), they show that shift and 
primary vector transformations each form a proper sub-algebra 
on their own,  and their algebras are Abelian.
Expression (\ref{PBb}) shows how the primary vector constraint 
changes under the shift transformation; indeed, it is observed 
that there is no substantial change since the vector constriction 
is still preserved but with a different test field.  
Relationships (\ref{PBc}) and (\ref{PBd}) reveal
how the shift transformations change under the scalar and vector constraints. At this point, the role played  by scalar and vector constraint as generators of diffeomorphisms, out and tangential, 
to $\Sigma_t$, is evident.  Likewise, (\ref{PBf}) and (\ref{PBg}), determine how the primary vector constraint changes under the 
scalar and vector constraints. To end this description, we mention 
that, despite the complete algebra is closed under the DB, this
is an open algebra since  several of the test fields, (\ref{test}),
depend on some of the phase space variables. Furthermore, this constraint algebra is not encountered in the usual gauge 
theories.  This fact   represents a difficulty  towards a standard  canonical quantization of GBG in the framework
considered.

%%%%%%%%%%%%%%%%%%%%%%%%%%%%%%%%%%%%%%%%%%%%%%%%%%%%%%%
\subsection{Infinitesimal canonical transformations}
%%%%%%%%%%%%%%%%%%%%%%%%%%%%%%%%%%%%%%%%%%%%%%%%%%%%%%%
In order to further illustrate the role of the constraints in the theory, in this subsection 
we consider infinitesimal canonical transformations.

It is worth remembering that, for any classical observable
$F \in \Gamma$, the Hamiltonian vector field
\be
X_F := \int_\Sigma
\left( \frac{\delta F}{\delta p} \cdot 
\frac{\delta}{\delta X} + 
\frac{\delta F}{\delta P} \cdot 
\frac{\delta}{\delta \dot{X}} - 
\frac{\delta F}{\delta X} \cdot 
\frac{\delta}{\delta p} - 
\frac{\delta F}{\delta \dot{X}} \cdot 
\frac{\delta}{\delta P} \right),
\ee
generates a one-parameter family of canonical transformations
$G \longrightarrow G + \delta_F G$, where $\delta_F G :=
\epsilon \{ G, F \}$, with $\epsilon$ being an infinitesimal dimensionless
quantity. The Hamiltonian vector fields associated with the 
first-class constraints (\ref{FC2}) induce the infinitesimal
canonical transformations
\be
\begin{array}{ll}
X_{\mathsf{F}_1} \longrightarrow 
\begin{cases}
\delta_{\mathcal{S}_\lambda} X^\mu = 0,
\\
\delta_{\mathcal{S}_\lambda} \dot{X}^\mu =
\epsilon_1 \lambda \dot{X}^\mu,
\\
\delta_{\mathcal{S}_\lambda} p_\mu = 0,
\\
\delta_{\mathcal{S}_\lambda} P_\mu = - \epsilon_1
\lambda P_\mu,
\end{cases}
& \qquad
X_{\mathsf{F}_3} \longrightarrow 
\begin{cases}
\delta_{S_\Lambda} X^\mu = \epsilon_3 \Lambda \dot{X}^\mu,
\\
\delta_{S_\Lambda} \dot{X}^\mu =  \epsilon_3 \frac{\delta S_\Lambda}{\delta P_\mu},
\\
\delta_{S_\Lambda} p_\mu = - \epsilon_3 \frac{\delta S_\Lambda}{\delta X^\mu} ,
\\
\delta_{S_\Lambda} P_\mu = - \epsilon_3 \frac{\delta 
S_\Lambda}{\delta \dot{X}^\mu},
\end{cases}
\\
X_{\mathsf{F}_2} \longrightarrow 
\begin{cases}
\delta_{\mathcal{V}_{\vec{\lambda}}} X^\mu = 0,
\\
\delta_{\mathcal{V}_{\vec{\lambda}}} \dot{X}^\mu =
\epsilon_2 \pounds_{\vec{\lambda}} X^\mu,
\\
\delta_{\mathcal{V}_{\vec{\lambda}}} p_\mu = \epsilon_2 
\pounds_{\vec{\lambda}} P_\mu,
\\
\delta_{\mathcal{V}_{\vec{\lambda}}} P_\mu = 0,
\end{cases}
& \qquad
X_{\mathsf{F}_4} \longrightarrow
\begin{cases}
\delta_{V_{\vec{\Lambda}}} X^\mu = \epsilon_4 \pounds_{\vec{\Lambda}}X^\mu,
\\
\delta_{V_{\vec{\Lambda}}} \dot{X}^\mu =
\epsilon_4 \pounds_{\vec{\Lambda}} \dot{X}^\mu,
\\
\delta_{V_{\vec{\Lambda}}} p_\mu = \epsilon_4 
\pounds_{\vec{\Lambda}} p_\mu,
\\
\delta_{V_{\vec{\Lambda}}} P_\mu = \epsilon_4 
\pounds_{\vec{\Lambda}} P_\mu,
\end{cases}
\end{array}
\label{CTs} 
\ee
where $\epsilon_u$, with $u=1,\ldots, 4$, denotes arbitrary 
gauge parameters corresponding to each of the first-class
constraints $\mathsf{F}_u$, respectively. 
For instance,
\be
\dot{X}^\mu \mapsto \dot{X}^\mu + \epsilon_1 \lambda \dot{X}^\mu ,
\qquad \text{and} \qquad
P_\mu \mapsto P_\mu - \epsilon_1 \lambda P_\mu,
\nonumber
\ee
are the gauge transformations induced by the gauge function
$\lambda$. From (\ref{CTs}) we infer that the constraint
$V_{\vec{\Lambda}}$ generates diffeomorphisms tangential to
$\Sigma_t$, while $S_\Lambda$ is the generator of diffeomorphisms
out of $\Sigma_t$ onto the worldvolume $m$. On the other hand,
$\mathcal{S}_\lambda$ is the generator of a 
%parity transformation (
momentum reflection %) 
in the sub-sector of $\Gamma$ given by $\{ 
\dot{X}^\mu; P_\mu \}$ that is, the sector associated to the 
second-order derivative dependence; from another view point, 
this constraint generates \textit{shift transformations} only in 
the velocity sector of the phase space. Finally, the constraint
$\mathcal{V}_{\vec{\lambda}}$ only acts on the sub-sector
$\{ \dot{X}^\mu; p_\mu \}$ by generating displacements in the 
orthogonal complement of this sub-sector, that is, in the 
sub-sector $\{ X^\mu; P_\mu \}$.

%%%%%%%%%%%%%%%%%%%%%%%%%%%%%%%%%%%%%%%%%%%%%%%%%%%%%%%%%%%%%%
\section{Discussion}
\label{sec6}
%%%%%%%%%%%%%%%%%%%%%%%%%%%%%%%%%%%%%%%%%%%%%%%%%%%%%%%%%%%%%%%

We have carried out a complete Ostrogradsky-Hamilton canonical study 
of  geodetic brane gravity described by the 
RT model in which the embedding functions of 
the brane are the field variables instead of the components of 
the metric, as in metric-GR.  An  essential ingredient in our analysis 
is the construction of an ADM Lagrangian density for the model linear in the embedding functions
acceleration. Usually this term is discarded as a boundary term contribution. By keeping it, we treat the
RT model as a higher derivative theory,  even though it has eom of second order. 
According to the Ostrogradsky-Hamilton canonical formulation, we have an extended phase space that
has positions and velocities as configuration canonical variables, together with their conjugate momenta.
We have derived the canonical Hamiltonian density for the model that contains terms linear in the conjugate momenta.
This is a signal of the well known Ostrogradsky instability for higher derivative theories, {\it i.e.} the Hamiltonian
is unbounded from below. However, one can hope that a suitable canonical transformation can be found 
to deal with this issue and obtain an Hamiltonian bounded from below. A possible strategy has been suggested
by Paul \cite{Paul2017}, and consists in solving the second class constraints, but it appears to be non trivial in the present case.
Another alternative is to implement a path integral quantization program adapted to second-order singular systems where second-class constraints are present in the theory \cite{Senjanovic1976} but, this deserves further investigation. 
As expected from the symmetry under reparametrization invariance of the theory,
the Hamiltonian is a linear combination of constraints. 
We have determined the complete set of constraints, 
and separated them into first- and second-class constraints. The appearance of second class constraints
is the price paid for keeping a term linear in the acceleration in the Lagrangian, however they take a form that
is quite manageable. In addition, we show explicitly how the constraints generate the expected gauge transformations, and
a correct counting of the physical degrees of freedom has 
been obtained.
We also checked that  Hamilton's equations reproduce the Euler-Lagrange equations of the theory. 
It should be mentioned that, based on the expressions for the Lagrangian and Hamiltonian densities as well from from the constraint densities obtained, the codimension is left arbitrary. Many of the features of the RT model generalize to the larger class of theories linear affine
in accelerations \cite{CMR2016}.

In principle, starting from our classical formulation, a  formal canonical quantization program can be
implemented. This would satisfy Regge and Teitelboim original motivation. With respect to quantum gravity one 
important technical advantage is the presence of a fixed background, that should come of help in a 
formal quantization. The phase space variables would be promoted to operators in a suitable Hilbert space.
As appropriate for a theory with second class constraints,
the Dirac brackets would turn into commutators for such operators \cite{HT1992,gitman1990}. 
A point to be confronted would be to find suitable gauge fixing conditions, to arrive at a space of physical states.
The issue of 
obtaining an Hamiltonian bounded from below, avoiding the presence of ghosts and lack of unitarity, would need
to be resolved.  In fact, ideally one would like to derive an Hamiltonian constraint quadratic in the momenta $p_\mu$. 
Another difficulty arises from the fact that the constraint algebra obtained is not a genuine Lie algebra, in addition to not being encountered in  usual gauge theories. All of these issues would also appear in a path integral or BRST quantization of the model.
Although aware of the difficulties ahead, we believe that our 
Ostrogradsky-Hamilton treatment of geodetic brane gravity
provides a reliable stepping stone.

%%%%%%%%%%%%%%%%%%%%%%%%%%%%%%%%%%%%%%%%%%%%%%%%%%%5
\section*{Acknowledgments}

GC acknowledges support from a CONACYT-M\'exico 
doctoral fellowship. ER acknowledges encouragment 
from ProDeP-M\'exico, CA-UV-320: \'Algebra, Geometr\'\i a 
y Gravitaci\'on. We are greatful to N. Kiriushcheva and
S. V. Kuzmin for providing with the preprint \cite{dutt1986}
and for useful comments. Also, RC and ER thanks partial support 
from Sistema Nacional de Investigadores, M\'exico.
%%%%%%%%%%%%%%%%%%%%%%%%%%%%%%%%%%%%%%%%%%%%%%%%%%%%%

%\section{Tables}

%\begin{table}[ph]
%\tbl{Comparison of acoustic for frequencies for piston-cylinder problem.}
%{\begin{tabular}{@{}cccc@{}} \toprule
%Piston mass & Analytical frequency & TRIA6-$S_1$ model &
%\% Error \\
%& (Rad/s) & (Rad/s) \\ \colrule
%1.0\hphantom{00} & \hphantom{0}281.0 & \hphantom{0}280.81 & 0.07 \\
%0.1\hphantom{00} & \hphantom{0}876.0 & \hphantom{0}875.74 & 0.03 \\
%0.01\hphantom{0} & 2441.0 & 2441.0\hphantom{0} & 0.0\hphantom{0} \\
%0.001 & 4130.0 & 4129.3\hphantom{0} & 0.16\\ \botrule
%\end{tabular} \label{ta1}}
%\end{table}

%If tables need to extend over to a second page, the continuation of
%the table should be preceded by a caption, e.g.~``{\it Table 2.}
%$(${\it Continued}$)$''.

\appendix

%%%%%%%%%%%%%%%%%%%%%%%%%%%%%%%%%%%%%%%%%%%%%%%%%%%%%%
\section{Integrability conditions}
\label{app0}
%%%%%%%%%%%%%%%%%%%%%%%%%%%%%%%%%%%%%%%%%%%%%%%%%%%%%%

Depending on the viewpoint, we have integrability 
conditions %useful 
to describe the geometry of an extended 
object at a fixed time, $\Sigma_t$, once this undergo an ADM 
split. 

%%%%%%%%%%%%%%%%%%%%%%%%%%%%%%%%%%%%%%%%%%%%%%%%%%%%
\subsection{$\Sigma_t$ embedded in $m$}
%%%%%%%%%%%%%%%%%%%%%%%%%%%%%%%%%%%%%%%%%%%%%%%%%%%%%

If $\Sigma_t$ is embedded into $m$, $x^a = \chi^a (u^A)$, with
$u^A$ being the local coordinates in $\Sigma_t$ and
$A = 1,2,\ldots,p$., the orthonormal basis is provided
by $\{ \epsilon^a{}_A = \partial_A \chi^a, \eta^a \}$. This 
satisfies $g_{ab} \epsilon^a{}_A \eta^b = 0$, $g_{ab}\eta^a \eta^b 
= -1$ and $g_{ab} \epsilon^a{}_A \epsilon^b{}_B = h_{AB}$
where $h_{AB}$ is the spacelike metric associated to $\Sigma$.
The corresponding Gauss-Weingarten (GW) equations are
\begin{equation}
\begin{aligned}
\nabla_A \epsilon^a{}_B & = \Gamma^C_{AB} \epsilon^a{}_C
+ k_{AB} \eta^a,
\\
\nabla_A \eta^a & = k_{AB} h^{BC} \epsilon^a{}_C,
\end{aligned}
\label{GW1}
\end{equation}
where $\nabla_A = \epsilon^a{}_A \nabla_a$, $k_{AB} = k_{BA}$
is the extrinsic curvature of $\Sigma_t$ associated to the normal
$\eta^a$ and $\Gamma^C_{AB}$ stands for the connection compatible
with $h_{AB}$.

\smallskip
\noindent
The intrinsic and extrinsic geometries for the embedding under
consideration must satisfy the integrability conditions
\begin{subequations}
\begin{align}
\mathcal{R}_{abcd} \epsilon^a{}_A \epsilon^b{}_B \epsilon^c{}_C
\epsilon^d{}_D &= R_{ABCD} - k_{AD}k_{BC} + k_{AC}k_{BD},
\\
\mathcal{R}_{abcd} \epsilon^a{}_A \epsilon^b{}_B \epsilon^c{}_C
\eta^d &= \D_A k_{BC} - \D_B k_{AC}, 
\end{align}
\end{subequations}
where $R_{ABCD}$ is the Riemann tensor associated to the spacelike
manifold $\Sigma_t$ and $\D_A$ is the covariant derivative compatible
with $h_{AB}$.

%%%%%%%%%%%%%%%%%%%%%%%%%%%%%%%%%%%%%%%%%%%%%%%%%%%%
\subsection{$\Sigma_t$ embedded in $\mathcal{M}$}
%%%%%%%%%%%%%%%%%%%%%%%%%%%%%%%%%%%%%%%%%%%%%%%%%%%%%

If $\Sigma_t$ is embedded into $\mathcal{M}$, $x^\mu = 
X^\mu (u^A)$, 
the orthonormal basis is provided by $\{ \epsilon^\mu{}_A = 
\partial_A X^\mu, \eta^\mu, n^\mu{}_i \}$. This 
satisfies $ \epsilon_A \cdot \eta = \epsilon_A \cdot n_i = \eta
\cdot n_i = 0$, $\eta \cdot \eta = -1$, $n_i \cdot n_j = \delta_{ij}$ 
and $ \epsilon_A  \cdot \epsilon_B = h_{AB}$.
The corresponding GW equations are
\begin{equation}
\begin{aligned}
D_A \epsilon^\mu{}_B & = \Gamma^C_{AB} \epsilon^\mu{}_C
+ k_{AB} \eta^\mu - L_A{}^i n^\mu{}_i,
\\
D_A \eta^\mu & = k_{AB} h^{BC} \epsilon^\mu{}_C - L_A{}^i
n^\mu{}_i,
\\
D_A n^{\mu\,i} &= L_{AB}^i h^{BC}\epsilon^\mu{}_C
- L_A{}^i \eta^\mu + \sigma_A{}^{ij} n^\mu{}_j,
\end{aligned}
\label{GW2}
\end{equation}
where $D_A = \epsilon^\mu{}_A D_\mu$ and $D_\mu$ being the 
background covariant derivative, $L_{AB}^i = L_{BA}^i$
is the extrinsic curvature of $\Sigma_t$ associated to the normal
$n^\mu{}_i$. 
Additionally, we have introduced $L_A{}^i := \epsilon^a{}_A 
\eta^b K_{ab}^i$ and $\sigma_A{}^{ij} := \epsilon^a{}_A 
\omega_a{}^{ij}$. Observe that the tangent-normal projection of the worldvolume
extrinsic curvature is in fact a piece of a non-trivial
twist potential given by $L_A{}^i = n^i \cdot D_A \eta$.

The intrinsic and extrinsic geometries for the embedding under
consideration must satisfy the integrability conditions
\begin{subequations}
\begin{align}
0 &= - R_{ABCD} - k_{AC}k_{BD} + k_{BC} k_{AD} + L_{AC}^i L_{BD\,i}
- L_{BC}^i L_{AD\,i},
\\
0 &= \D_A k_{BC} - \D_B k_{AC} + L_A{}^i L_{BC\,i} - L_B{}^iL_{AC\,i},
\\
0 & = \Dt_A L_{BC}^i - \Dt_B L_{AC}^i + L_A{}^i k_{BC} - L_B{}^i
k_{AC},
\\
0 & = \Dt_A L_B{}^i - \Dt_B L_A{}^i + L_A{}^{C\,i} k_{BC}
- L_B{}^{C\,i}k_{AC},
\\
0 & = - \Omega_{AB}^{ij} + L_A{}^{C\,i} L_{BC}^j - L_B{}^{C\,i}
L_{AC}^j - L_A{}^i L_B{}^j + L_B{}^i L_A{}^j,
\end{align}
\end{subequations}
where $\Omega_{AB}^{ij} : = \Dt_B \sigma_A{}^{ij} - 
\Dt_A \sigma_B{}^{ij}$ is the curvature tensor associated with 
the gauge field $\sigma_A{}^{ij}$ and $\Dt_A$ is the $O(N - p - 2)$ 
covariant derivative acting on the normal indices associated
with the connection $\sigma_A{}^{ij}$.

%%%%%%%%%%%%%%%%%%%%%%%%%%%%%%%%%%%%%%%%%%%%%%%%
\section{Functional derivatives of $S_\Lambda$}
\label{app1}
%%%%%%%%%%%%%%%%%%%%%%%%%%%%%%%%%%%%%%%%%%%%%%%%%%%

Here we present the functional derivatives of the 
Hamiltonian constraint $S_\Lambda$
{\footnotesize
\begin{equation}
\begin{aligned} 
\frac{\delta S_\Lambda}{\delta p_\mu},
&=
\Lambda\,\dot{X}^\mu
\qquad \qquad \qquad \qquad \qquad
\frac{\delta S_\Lambda}{\delta P_\mu}
=
2\Lambda N^A\D_A \dot{X}^\mu 
+ \Lambda (N^2 h^{AB} - N^A N^B) \D_A \D_B X^\mu,
\\
\frac{\delta S_\Lambda}{\delta \dot{X}^\mu}
&=
\Lambda\,p_\mu 
+ 2\Lambda (P \cdot \D_A \dot{X})h^{AB} \epsilon_{\mu\,B} 
- \D_A (2\Lambda N^A\,P_\mu)
\\
&- 
2\Lambda N h^{AB} (P \cdot \D_A \D_B X)\, \eta_\mu
-2 \Lambda N^B (P \cdot \D_A \D_B X) h^{AC}\epsilon_{\mu C}
\\
& - \frac{1}{2}\sqrt{h}\Lambda G^{ABCD} \Pi_{\alpha\beta}
\mathcal{D}_A \mathcal{D}_B X^\alpha \mathcal{D}_C \mathcal{D}_D
X^\beta\,\eta_\mu
+  \Lambda \sqrt{h}\,G^{ABCD}L_{AB}^i k_{CD}\,n_{\mu\,i}
\\
&-  \Lambda \sqrt{h} h^{AB} \delta_{ij}\, 
\left( \Dt_A n^i \cdot \Dt_B n^j \right)\eta_\mu
-  \left[ 
\Dt_A \left( 2 \Lambda N \sqrt{h} h^{AB} \delta_{ij}
\Dt_B n^j \right) \cdot \eta \frac{1}{N} \right] n_\mu{}^i,
\\
\frac{\delta S_\Lambda}{\delta X^\mu}
&= \D_A \widetilde{T}^A,
\end{aligned}
\end{equation}
where
\beq
\widetilde{T}^A &:=&  - 2\Lambda N h^{AB}(P\cdot \D_B \dot{X})\,
\eta_\mu + 2\Lambda N^A h^{BC} 
(P \cdot \D_B \dot{X}) \,\epsilon_{\mu\,C}
- 2\Lambda N^A h^{BC}N (P \cdot \D_B \D_C X)
\,\eta_\mu
\nonumber
\\
&+& 2\Lambda N^C h^{AB} N (P \cdot \D_B \D_C X)
\,\eta_\mu - 2\Lambda N^B h^{CD} N^A (P \cdot \D_B 
\D_C X)\,\epsilon_{\mu\,D}
\nonumber
\\
&+&  2\Lambda N^2h^{AC}h^{BD} (P \cdot \D_B \D_C X)
\,\epsilon_{\mu \,D}
+ \D_B \left[ \Lambda (N^2 h^{AB} - N^A N^B)
\,P_\mu \right]
\nonumber
\\
&-& \frac{1}{2}\Lambda 
\sqrt{h} N^A G^{BCDE} \Pi_{\alpha\beta} \mathcal{D}_B \mathcal{D}_C X^\alpha \mathcal{D}_D \mathcal{D}_E X^\beta \,\eta_\mu 
\nonumber
\\
&-& \frac{1}{2}\Lambda N 
\sqrt{h} G^{CDEF} \Pi_{\alpha\beta} \mathcal{D}_C \mathcal{D}_D X^\alpha \mathcal{D}_E \mathcal{D}_F X^\beta \,h^{AB}\epsilon_{\mu\,B}
+  2  \Lambda N\sqrt{h} (L^{AB}_i L^i
- L^{AC}_i L_C{}^{B\,i})\epsilon_{\mu B}
\nonumber
\\
&+& \Lambda \sqrt{h}N^A G^{BCDE}L_{BC}^i k_{DE}
\,n_{\mu\,i}
+ \mathcal{D}_B ( \Lambda N \sqrt{h}
G^{ABCD} \Pi_{\mu \nu} \mathcal{D}_C \mathcal{D}_D X^\nu)
\nonumber
\\
&-& \Lambda \sqrt{h} h^{BC} \delta_{ij}\, 
\left( \Dt_B n^i \cdot \Dt_C n^j \right) N^A\, \eta_\mu
- \Lambda N\sqrt{h}h^{AB}  h^{CD} 
\delta_{ij} \left(\Dt_C n^i \cdot \Dt_D n^j\right)\epsilon_{\mu B}
\nonumber
\\
&+& 2\Lambda N \sqrt{h} h^{AB} h^{CD} 
\delta_{ij} \left( \Dt_B n^i \cdot \Dt_C n^j\right) \epsilon_{\mu \,D}
- \left[ 
\Dt_C \left( 2 \Lambda N \sqrt{h} h^{CD} \delta_{ij}
\Dt_D n^j \right) \cdot \epsilon_B\,h^{AB}
 \right]n_\mu{}^i
\nonumber
\\
&-& \left[ 
\Dt_C \left( 2 \Lambda N \sqrt{h} h^{BC} \delta_{ij}
\Dt_B n^j \right) \cdot \eta\,\frac{N^A}{N}
 \right]n_\mu{}^i.
 \label{FDs3} 
\eeq}
is a vector field with density weight $+1$.

%%%%%%%%%%%%%%%%%%%%%%%%%%%%%%%%%%%%%%%%%%%%%%%%
\section{Constraint algebra}
\label{app2}
%%%%%%%%%%%%%%%%%%%%%%%%%%%%%%%%%%%%%%%%%%%%%%%%%%%

Primary-primary constraints 
\begin{equation}
\begin{array}{ll}
\{ \mathcal{S}_\lambda , \mathcal{S}_{\lambda'} \} = 0,
& \qquad \qquad \qquad
\{ \mathcal{V}_{\vec{\lambda}}, \mathcal{V}_{\vec{\lambda}'} \} = 0,
\\
\{ \mathcal{S}_\lambda, \mathcal{V}_{\vec{\lambda}} \}
= \mathcal{V}_{\vec{\lambda}'} \qquad \,\,\,\,\lambda^{'A} = \lambda \lambda^A,
& \qquad \qquad \qquad
\{ \mathcal{V}_{\vec{\lambda}}, \mathcal{W}_{\vec{\phi}} \}
= 0,
\\
\{ \mathcal{S}_\lambda , \mathcal{W}_{\vec{\phi}} \} = 
\mathcal{W}_{\vec{\phi}^{\,'}} \qquad
\phi^{'i} = \lambda \phi^i,
& \qquad \qquad \qquad 
\{ \mathcal{W}_{\vec{\phi}}, \mathcal{W}_{\vec{\phi}^{\,'}} \}
= 0,
\end{array}
\label{algebra1A}
\end{equation}
Primary-secondary constraints 
\begin{equation}
\begin{array}{ll}
\{ \mathcal{S}_\lambda , S_\Lambda \} = - \mathcal{S}_{\lambda_1}
- S_{\Lambda_1},
& \qquad
\{ \mathcal{W}_{\vec{\phi}} , S_\Lambda \} = \mathcal{S}_{\lambda_2}
+ \mathcal{V}_{\vec{\lambda}_2} - \mathcal{W}_{\vec{\phi}_1}
- W_{\vec{\Phi}_1},
\\
\{ \mathcal{S}_\lambda , V_{\vec{\Lambda}} \}
= - \mathcal{S}_{\pounds_{\vec{\Lambda}}\lambda},
& \qquad
\{ \mathcal{W}_{\vec{\phi}} , V_{\vec{\Lambda}} \} = 
\mathcal{S}_{\lambda_3} - \mathcal{V}_{\vec{\lambda}_3}
- \mathcal{W}_{\vec{\phi}_2},
\\
\{ \mathcal{S}_\lambda , W_{\vec{\Phi}} \} = \mathcal{A},
& \qquad
\{ \mathcal{W}_{\vec{\phi}} , W_{\vec{\Phi}} \} = \mathcal{W}_{\vec{\phi}_3} + \mathcal{C},
\\
\{ \mathcal{V}_{\vec{\lambda}} , S_\Lambda \} = 
\mathcal{S}_{\pounds_{\vec{\lambda}}\Lambda} 
- \mathcal{V}_{\vec{\lambda}_1} - V_{\vec{\Lambda}_1},
& \qquad
\\
\{ \mathcal{V}_{\vec{\lambda}} , V_{\vec{\Lambda}} \} = \mathcal{V}_{[\vec{\lambda}, \vec{\Lambda}]},
& \qquad
\\
\{ \mathcal{V}_{\vec{\lambda}} , W_{\vec{\Phi}} \} = \mathcal{B},
& \qquad
\end{array}
\label{algebra2A}
\end{equation}
where
\begin{equation}
\begin{array}{ll}
\lambda_1 = 2\Lambda \pounds_{\vec{N}} \lambda,
& \quad 
\phi^i_1 = \Lambda \phi^i N k - \phi^i N^A \mathcal{D}_A \Lambda
+ \Lambda N^A \Dt_A \phi^i,
\\
\Lambda_1 = \lambda \Lambda,
& \quad
\Phi^i_1 = \Lambda \phi^i,
\\
\lambda^A_1 = 2 \Lambda N^B \mathcal{D}_B \lambda^A,
& \quad
\lambda_3 = \frac{\Lambda^A}{N} \phi_i L_A{}^i,
\\
\Lambda^A_1 = \Lambda \lambda^A,
& \quad
\lambda^A_3 = \frac{N^A}{N} \phi_i \Lambda^B L_B{}^i + \phi_i \Lambda^B L_B{}^{A\,i},
\\
\lambda_2^A = \Lambda \phi_i \left( N^A L^i - N^B L_{B}{}^{A\,i}
- \frac{N^A N^B}{N} L_B{}^i \right),
& \quad
\phi^i_2 = \Lambda^A \Dt_A \phi^i,
\\
\lambda_2^A = \Lambda \phi_i \left( N^A L^i - N^B L_{B}{}^{A\,i}
- \frac{N^A N^B}{N} L_B{}^i \right),
& \quad
\phi^i_3 =  \frac{N^A}{N}\phi^i \Phi_j\,L_A{}^i,
\end{array}
 \nonumber
\end{equation}
and {\footnotesize
\begin{eqnarray}
\mathcal{A} &=& \int_\Sigma \lambda \Dt_A \Phi^i
 \sqrt{h} h^{AB} (n_i \cdot \D_B \dot{X}),
\\
\mathcal{B} &=& - \int_\Sigma  \sqrt{h} h^{AB}
\lambda^C \Dt_A \Phi_i \,L_{BC}^i,
\\
\mathcal{C} &=& \int_\Sigma \left[ 
 \frac{\sqrt{h}}{N^2} \phi^i\Phi_i\, L_A{}^j
(N^A L_j + N L^A{}_j) - \frac{1}{N}\phi^i \Phi_i 
(p \cdot \eta) + \frac{\sqrt{h}}{N} \phi^i \Phi_j
(L^{AB}_i L_{AB}^j - L_i L^j) \right]
%\right. 
%\nonumber
\end{eqnarray}}
Secondary-secondary constraints
\begin{equation}
\begin{array}{ll}
\{ S_\Lambda , S_{\Lambda'} \} = \mathcal{S}_{\lambda_4} +
\mathcal{W}_{\vec{\phi}_4},
& \qquad \qquad \qquad
\{ V_{\vec{\Lambda}}, V_{\vec{\Lambda}^{\,'}} \} = V_{[\vec{\Lambda},
\vec{\Lambda}^{\,'}]},
\\
\{ S_\Lambda, V_{\vec{\Lambda}} \}
= \mathcal{V}_{\vec{\lambda}_4} 
- \mathcal{S}_{\pounds_{\vec{\Lambda}}\Lambda},
& \qquad \qquad \qquad
\{ V_{\vec{\Lambda}}, W_{\vec{\Phi}} \}
= \mathcal{S}_{\lambda_5} + \mathcal{V}_{\vec{\lambda}_5}
+ \mathcal{W}_{\vec{\phi}_4} + W_{\vec{\Phi}_4}
+ \mathcal{E},
\\
\{ S_\Lambda , W_{\vec{\Phi}} \} = \mathcal{S}_{\lambda_6}+\mathcal{V}_{\vec{\lambda}_6}+\mathcal{D},
& \qquad \qquad \qquad 
\{ W_{\vec{\Phi}}, W_{\vec{\Phi}^{\,'}} \}
= - \mathcal{W}_{\vec{\phi}_5} + \mathcal{F},
\end{array}
\label{algebra3A}
\end{equation}
where
\begin{eqnarray}
\lambda_4 &=& (N^2 h^{AB} - N^A N^B) (\Lambda \D_A \D_B \Lambda'
- \Lambda' \D_A \D_B \Lambda),
\\
\phi^i_4 &=& 2N^3h^{AB} L_A{}^i (\Lambda \D_B \Lambda' - \Lambda'
\D_B \Lambda),
\\
\lambda_4^A &=& \Lambda (N^2 h^{BC} - N^B N^C)(\D_B \D_C \Lambda^A 
+ R_{BDC}{}^A \Lambda^D),
\\
\lambda_5 &=& \frac{1}{N} \left[ \Lambda^A N^B k_A{}^C L_{BC}^i \,\Phi_i 
- \pounds_{\vec{\Lambda}} (N^A L_A{}^i)\Phi_i - \Lambda^A L_A{}^i
\pounds_{\vec{N}} \Phi_i
\right],
\\
\lambda_5^C &=& \Lambda^A \Dt_A (N^B L_B{}^{C\,i})\Phi_i + \Lambda^A \Dt_A (N^B L_B{}^i)
\frac{N^C}{N}\Phi_i
\n
\\
&+& \Lambda^A \frac{N^B}{N} \left[ N\Dt_B \Phi_i \,L_A{}^{C\,i} - k_A{}^D L_{BD}^i N^C 
\Phi_i + \frac{N^B}{N} (L_A{}^i \Dt_B \Phi_i \,N^C 
\right.
\n
\\
&-& \left. N k_A{}^C L_B{}^i \,\Phi_i)\right],
\\
\phi_4^i &=& 2 \Lambda^{[A}N^{B]} \left( L_A{}^{C\,j}L_{BC}^i 
- L_A{}^j L_B{}^i\right)\,\Phi_j,
\\
\Phi_4^i &=& \Lambda^A \Dt_A \Phi^i,
\\
\phi_5^i &=& N h^{AB} K_B{}^j ( \Phi_j \Dt_A \Phi^{'\,i}
- \Phi'_j \Dt_A \Phi^i ),
\\
\lambda_6^C&=&\Lambda \Phi_i\left[ \left( N^2 h^{AB}
+ N^A N^B \right) \left( \Dt_A L^{iC}{}_B-L^i{}_B 
k_A{}^C \right) + NN^C\Dt_AL^{A\,i} 
\right.  
\n
\\
&+&\left.  2 N^A L_B{}^{C\,i} \D_A N^B 
+ 2\frac{N^AN^C}{N}L^i{}_B\D_AN^B 
+ \frac{N^AN^BN^C}{N}	\Dt_AL^i{}_B \right] 
\n
\\
&+&\Lambda\Dt_A\Phi_i\left[ 2\left( h^{AB}N^2+N^AN^B
\right)\left( L^{iC}{}_B+\frac{L^i{}_BN^C}{N}\right)
\right],
\\
\lambda_6 &=&-\Lambda\Phi_i\left[\left( h^{AB}N^2+N^AN^B\right)\frac{\Dt_A L^i{}_B}{N}+2N^AL^i{}_B\D_AN^B\right]
\\
&+& \Lambda\Dt_A\Phi_i\left[ -2\left( h^{AB}N^2+N^AN^B\right)\frac{L^i{}_B}{N}\right],
\end{eqnarray} 
and
\begin{equation}
\begin{split}
\mathcal{D} &=\int_\Sigma \left\lbrace \Lambda \Phi_i
\left[ 2 N L_A{}^i\left( P\cdot \D^A\dot{X} \right)- 2N 
\left(  N^BL^{A\,i}+ NL^{AB\,i}\right)
\left( P\cdot \D_A\D_B X\right)
\right.\right.
\\   
&-\left[ N^2\left( L^i{}_AL^{jA}-L^i{}_{AB}L^{jAB}\right) -NN^A\left( L^iL^j{}_A-L^i{}_BL^{jB}{}_A\right) \right]\left( P\cdot n_j\right)
\\
&+\sqrt{h}\left[\frac{N}{2}L^iL^jL_j+\frac{N}{2}L^iL^{jCD}L_{jCD}-NL^jL_j{}^{AB}L^i{}_{AB}-NL_j{}^A L^i{}_{AD}L^{jD} 
\right.
\\
&\left.\left. - N L^{AC}_j L_C{}^{B\,j} L_{AB}^i + 4 N L_{jA}L^{(i}{}_BL^{j)AB}+2N^CL_j{}^AL^{[i}{}_AL^{j]}{}_C - 2NL^iL_j{}^AL^j{}_A\right]\right]
\\
&+\Lambda\Dt_A\Phi_i\left[ N^A\left( p\cdot n^i\right) -\left( P\cdot n^i\right)\left( N^A\D_CN^C+N\D^AN\right)-\Dt_B\left( P\cdot n^i\right)\left( h^{AB}N^2+N^AN^B\right)  \right.
\\
&\left.+\sqrt{h}N\left(\left( k^{AB}-h^{AB}k\right)L^i{}_B+\Dt_BL^{iAB}\right)+2 \sqrt{h}N^{[B}\Dt_BL^{\vert i\vert A]}- \sqrt{h}N^C\Dt^AL^i{}_C\right]
\\
&\left.+\Lambda\Dt_A\Dt_B\Phi_i\left( 2\sqrt{h}L^{i[B}N^{A]}\right)\right\rbrace,
\end{split}
\end{equation}
\begin{eqnarray}
\mathcal{E} &=& \int_\Sigma \sqrt{h} \left\lbrace 
\frac{2}{N}\Lambda^{[A}N^{B]} (L_A{}^{C\,i}L_{BC}^j - L_A{}^i L_B{}^j)\, L_j \Phi_i - 2 \Lambda^C L_C{}^{[i} L_A{}^{j]}  L^A{}_j \Phi_i
\right.
\nonumber 
\\
&+& \left.  2h^{A[B}h^{D]C} \D_B \Lambda_C\,L_D{}^i \Dt_A \Phi_i
+ \Lambda^D (L_{CD}^i k^{CA} - k_{DC}L^{CA\,i})\Dt_A \Phi_i 
\right. 
\nonumber
\\
&+& \left. 2 \Lambda^D  L^A{}_j h^{BC} L_{BD}^{[i} L_{AC}^{j]} \Phi_i
 + 2 h^{A[B} h^{D]C} \Lambda_C L_D{}^i \Dt_A \Dt_B \Phi_i 
\right\rbrace,
\\
\mathcal{F} &=& \int_\Sigma  \left\lbrace \left[ 
- h^{AB} (p \cdot \partial_A X) + N^A L_{AC}^l h^{CB} (P \cdot n_l)
+  \sqrt{h}\,L_A{}^l L^{AB}_l
\right] \times\right.
\nonumber
\\
&& \left. \quad \qquad \qquad \qquad \qquad \qquad \qquad 
\qquad \qquad \qquad 
\times \delta_{ij}\left( \Phi^i \Dt_B \Phi^{'\,j}
- \Phi^{'\,i} \Dt_B \Phi^j \right)
\right.
\nonumber
\\
&-& \left. 2\sqrt{h}\,L_{A\,(i} \left( h^{AB}L_{j)} - L^{AB}_{j)} \right)
\left( \Phi^i \Dt_B \Phi^{'\,j} - \Phi^{'\,i}\Dt_B \phi^j \right)
\right\rbrace.
\end{eqnarray}

%\newpage

%\begin{thebibliography}{000} %for 3 digits
%\begin{thebibliography}{00}  %for 2 digits

\end{document}